\documentclass[fleqn,usenatbib]{mnras}

\usepackage[T1]{fontenc}
\usepackage{orcidlink}
\DeclareRobustCommand{\VAN}[3]{#2}
\let\VANthebibliography\thebibliography
\def\thebibliography{\DeclareRobustCommand{\VAN}[3]{##3}\VANthebibliography}

\usepackage{graphicx}	% Including figure files
\graphicspath{{./}}
\usepackage{amsmath}	% Advanced maths commands
\usepackage{amssymb}	% Extra maths symbols

\newcommand{\msun}     {\ensuremath{{M}_{\scriptscriptstyle \odot}}}

\newcommand{\units}[1]  {\ensuremath{\mathrm{{#1}}}}
\newcommand{\kgbf}[1]{{{#1}}}

\title[IMBHs on the FP]{Intermediate-mass black holes and the fundamental plane of black hole accretion}

\author[K.\ G{\"u}ltekin et al.]{
Kayhan G{\"u}ltekin$^{1}$\,\orcidlink{0000-0002-1146-0198}\thanks{Corresponding author (kayhan@umich.edu)}
Kristina Nyland$^{2}$\,\orcidlink{0000-0003-1991-370X},
Nichole Gray$^{1}$,
Greg Fehmer$^{1}$,
Tianchi Huang$^{1}$,
\newauthor Matthew Sparkman$^{1}$,
Amy E. Reines$^{3}$\,\orcidlink{0000-0001-7158-614X}, 
Jenny E. Greene$^{4}$\,\orcidlink{0000-0002-5612-3427}, 
Edward M. Cackett$^{5}$\,\orcidlink{0000-0002-8294-9281},
and Vivienne Baldassare$^{6}$\,\orcidlink{0000-0003-4703-7276}
\\
$^{1}$Department of Astronomy, University of Michigan, 1085 S.\ University Ave., Ann Arbor, MI 48104, USA\\
$^{2}$U.S. Naval Research Laboratory, 4555 Overlook Ave SW, Washington, DC 20375, USA\\
$^{3}$eXtreme Gravity Institute, Department of Physics, Montana State University, Bozeman, MT 59717, USA \\
$^{4}$Department of Astrophysical Sciences, Princeton University, Princeton, NJ 08544, USA \\
$^{5}$Department of Physics and Astronomy, Wayne State University, 666 W Hancock St, Detroit, MI 48201, USA\\
$^{6}$Department of Physics and Astronomy, Washington State University, 100 Dairy Road, Pullman, WA 99163, USA\\
}

\date{Accepted 2022 September 09. Received 2022 September 09; in original form 2022 April 09}

\pubyear{2022}
\usepackage{newtxtext,newtxmath}

\begin{document}
\label{firstpage}
\pagerange{\pageref{firstpage}--\pageref{lastpage}}
\maketitle

\begin{abstract}
We present new 5 GHz VLA observations of a sample of 8 active intermediate-mass black holes with masses $10^{4.9} < M < 10^{6.1}\ \msun$ found in galaxies with stellar masses $M_{*} < 3 \times 10^{9}\ \msun$.  We detected 5 of the 8 sources at high significance.  Of the detections, 4 were consistent with a point source, and one (SDSS J095418.15+471725.1, with black hole mass $M < 10^{5}\ \msun$) clearly shows extended emission that \kgbf{has a jet morphology}.  Combining our new radio data with the black hole masses and literature X-ray measurements, we put the sources on the fundamental plane of black hole accretion.  We find that the extent to which the sources agree with the fundamental plane depends on their star-forming/composite/AGN classification based on optical narrow emission line ratios.  The single star-forming source is inconsistent with the fundamental plane.  The three composite sources are consistent, and three of the four AGN sources are inconsistent with the fundamental plane.  We argue that this inconsistency is genuine and not a result of misattributing star-formation to black hole activity.  \kgbf{Instead, we identify the sources in our sample that have AGN-like optical emission line ratios as not following the fundamental plane and thus caution the use of the fundamental plane to estimate masses without additional constraints, such as radio spectral index, radiative efficiency, or the Eddington fraction.}
\end{abstract}

\begin{keywords}
quasars: supermassive black holes --- galaxies: active --- radio continuum: galaxies --- X-rays: galaxies
\end{keywords}

%%%%%%%%%%%%%%%%%%%%%%%%%%%%%%%%%%%%%%%%%%%%%%%%%%

%%%%%%%%%%%%%%%%% BODY OF PAPER %%%%%%%%%%%%%%%%%%

\section{Introduction}
\label{intro}
Intermediate-mass black holes (IMBHs) are black holes with masses in the range $10^2 < M / \msun < 10^6$.  This mass range is above the limit that can be produced from normal, isolated stellar evolution \citep[e.g.,][]{2006ARA&A..44...49R}, and below the masses more frequently found at the centers of galactic nuclei \citep[e.g.,][]{2013ARA&A..51..511K}.  Black holes have been found to exist just above the $10^2\ \msun$ limit via gravitational wave detection and show that stellar mass black holes (often called stellar-origin black holes in the gravitational wave community) can grow to intermediate masses \citep{2020PhRvL.125j1102A, 2020ApJ...900L..13A}.  On the other end of the intermediate mass range, many IMBHs and IMBH candidates are seen at the centers of galactic nuclei \citep{2020ARA&A..58..257G} including dwarf galaxies \citep{Reines2022}.  

IMBHs as the smallest central black holes are interesting for our understanding the growth and evolution of all central black holes.  Since most black holes at the current epoch are much larger than the masses they started with, IMBHs may be closer to the initial masses of all supermassive black holes \citep{2001ApJ...551L..27M, 2001ApJ...550..372F}.  IMBHs merging with other IMBHs or smaller objects will also emit gravitational wave at a frequency that Laser Interferometer Space Antenna (LISA) will be most sensitive to \citep{2017arXiv170200786A}.  The rates of such gravitational wave events is poorly predicted because the number density of IMBHs is poorly known \citep{2016ApJ...819....3M}.  IMBHs may also play a role in dense stellar clusters and are likely essential players in feedback for the smallest galaxies \citep{2022Natur.601..329S} and possibly the reionization of the early universe \citep{2002MNRAS.330..232C, 2006ApJ...640..156G, 2004ApJ...616..221G, 2016ASSL..418..317B, 2019ApJ...884..180D, 2015ApJ...813L...8M}.  IMBHs are also sources of tidal disruption events, which are and will be identified with time-domain surveys \citep{2009MNRAS.400.2070S}.  See \citet{2020ARA&A..58..257G} for a recent review on IMBHs.

Merely finding IMBHs moves these fields forward, but finding IMBHs requires knowing their mass, and measuring black hole masses is not trivial.  Dynamical mass measurements of black holes requires good angular resolution in order to isolate the gravitational effects of the black hole from the rest of the galaxy \citep[e.g.,][]{2009ApJ...698..198G}.  Smaller black holes have a smaller effect and thus require better angular resolution and/or smaller distances to them in order to measure their masses \citep[e.g.,][]{2010ApJ...714..713S, 2015ApJ...809..101D, 2017ApJ...836..237N, 2018ApJ...858..118N, 2020MNRAS.496.4061D}.  So most IMBHs and IMBH candidates have been identified via active galactic nuclei (AGN) line measurements \cite[e.g.,][]{2013ApJ...775..116R, 2015ApJ...809L..14B}.  These so-called single-epoch or virial mass measurements are widely used for the more massive black holes, but at low masses additional challenges are present \citep{2014SSRv..183..253P, 2014ApJ...789...17H}.  For example, smaller black holes will have lower-velocity-width broad lines, making them harder to distinguish from, e.g., a supernova without multi-epoch observations \citep{2016ApJ...829...57B}.  
%% Second, calibration of the single-epoch mass measurement method comes from a population of predominantly larger black holes.  So extrapolating the relation to the smallest masses may not be valid.  Finally, the smallest black holes are likely to come disproportionately from galaxies with pseudobulges, and the black-hole scaling relations that underpin the single-epoch methods do not apply for pseudobulges.  
Thus there is reason to look for as much corroborating evidence of an IMBH as possible.

One potential way to corroborate a black hole's mass is to use the fundamental plane of black hole accretion.  This fundamental plane is an empirical correlation between an accreting black hole's mass ($M$), radio luminosity ($L_R$), and X-ray luminosity  \citep[$L_X$;][]{2003MNRAS.345L..19M, 2003MNRAS.345.1057M, 2004A&A...414..895F, 2006ApJ...645..890W, 2006NewA...11..567M,  2008ApJ...688..826L,  2009ApJ...706..404G,  2012MNRAS.419..267P, 2014ApJ...787L..20D, 2014ApJ...788L..22G, 2016ApJ...818..185F,  2017ApJ...836..104X,  2018ApJ...860..134Q, 2018ApJ...862...16T, 2020ApJ...900..134W}.  There is strong observational evidence that it holds for stellar-mass black holes and central black holes across a wide range of Eddington ratios \citep[$f_{\mathrm{Edd}} \sim 10^{-9}$--$10^{-2}$;][]{2003MNRAS.344...60G, 2012MNRAS.423..590G, 2013ApJ...773...59P, 2019ApJ...871...80G}.  
The physics of the fundamental plane is incompletely understood but the fact that it applies to a wide range of black hole masses suggests that \kgbf{radiatively inefficiently accreting black holes have a common regulation of energy between X-ray emission (which is produced either via syncrotron emission or coronal emission after inverse Compton scattering) and radio emission (which arises directly from jet emission) \citep{2003MNRAS.343L..59H}.  In contrast to the radiatively inefficient accretion mechanisms thought to be operating in the fundamental plane, radiatively efficient accretion can produce X-rays directly from thermal thin-disk emission.}

Different physical interpretations of the empirical fundamental plane lead to 
different predictions when extrapolating to Eddington ratios or masses not well represented in the current observational data.  By adding data from these extreme Eddington ratios and/or mass, we can test the physical interpretations.  
Thus by studying IMBHs with the fundamental plane, we can determine the limits of the plane as an empirical correlation and therefore its usefulness as an additional way to estimate a black hole's mass.  If the correlation holds over the broadest set of black holes, then it can be used, e.g., to disambiguate between IMBHs and larger central black holes.

In this paper, we study a sample of 8 IMBHs with masses in the range $10^5 < M / \msun < 10^6$ based on single-epoch mass estimates \citep{2013ApJ...775..116R}.  Chandra X-ray Observatory observations showed these sources all to have X-ray point sources at their centers \citep{2017ApJ...836...20B}.  In this paper we present new 5 GHz radio observations of these sources to test whether they lie on the fundamental plane of black hole accretion.  If they do lie on the plane, then this gives confidence for the fundamental plane as an identification technique for sources similar to them. 

In section \ref{obs} we describe selection of the sample and our new radio observations of them.   We put together archival data and our new data to test the fundamental plane in section \ref{discuss}, and we summarize our findings in section \ref{concl}.

\section{Sample and Data}
\label{obs}
To test whether IMBH AGN are on the fundamental plane, we need a sample of IMBH AGN with X-ray and radio data. 
In this section we describe the creation of our sample of IMBH AGN, the literature X-ray observations of them, and our new radio observations of them.

\subsection{Sample creation and X-ray flux measurements}
We constructed a sample based on the optical AGN emission line identification of IMBH AGN.
These AGN were identified by \citet{2013ApJ...775..116R} based on Sloan Digital Sky Survey (SDSS) data by looking at galaxies with stellar masses of $M_{*} < 3 \times 10^{9}\ \msun$.  They fit the SDSS spectra and examined standard narrow emission line ratios for indications of an AGN as the source of the ionization---the \citet{2006MNRAS.372..961K} version of \citet[BPT]{1981PASP...93....5B}.  Some of the sources identified as AGN or as composite AGN/starforming also have broad H$\alpha$ emission from which black hole masses were estimated, the vast majority of which were smaller than $10^6\ \msun$.  Further time-domain spectroscopic analysis  showed that broad H$\alpha$ in BPT star-forming galaxies faded and was likely due to transient stellar processes (e.g., supernovae) rather than AGNs 
\citep{2016ApJ...829...57B}.

A follow-up study with the Chandra X-ray Observatory by \citet{2017ApJ...836...20B} showed that eight of the broad-H$\alpha$ galaxies (4 AGN, 3 composite, 1 star-forming) had point-source X-ray emission at the nucleus, a strong indication of black hole activity.  While all sources were detected in X-ray full band (0.3--7 keV), one of them was undetected in the hard band (2--7 keV) used for fundamental plane analysis.  We used these 8 sources as the basis of our sample for observation with the Karl G.\ Jansky Very Large Array (VLA).  The sources are listed in Table \ref{tab:vladata} along with their luminosity distances.

% We used NSF's KGJ VLA (VLA).
% Observation details: frequency, array config, exposure time.  
% Reduction details: pipeline, self-cal, cleaning, weights.
% Final imaging and flux analysis.

% Notes on individual sources: extended source, resolved, partially resolved, unresolved (with sizes).

\subsection{New radio observations}
\label{data}
We used the VLA to measure the radio continuum properties of our sample at $C$ band (4--8~GHz).  The observations were taken between 2015 June 26 and 2015 July 30 during the VLA A configuration (Project 15A-240; PI: K.\ G{\"u}ltekin), providing a typical angular resolution of $\sim$0.3$^{\prime \prime}$.  Each source was observed for 3 hours in one scheduling block, except for SDSS J095418.15+471725.1, which was observed for 4.5 hours split evenly into two scheduling blocks.  All scheduling blocks began with observations of a flux calibrator and then proceeded with iterating between observing a complex gain calibrator and the target source.  The iterations always started and ended with a calibration integration with typical integration times of 1 min for calibrator and 5 min for target source.  Typical total time on source for the target source was 120 minutes.

The data were processed using the scripted VLA 
calibration pipeline\footnote{\url{https://science.nrao.edu/facilities/vla/data-processing/pipeline/scripted-pipeline}} for the Common Astronomy Software Applications package \citep{2007ASPC..376..127M} version 5.3.0.  Imaging was performed using the TCLEAN task in CASA with standard parameters appropriate for broadband data with point-source emissison.  We used Briggs weighting with robust parameters between 0 and 1. Self calibration was implemented manually on an as-needed basis for sufficiently bright sources with evidence for residual phase errors in the image plane.  The typical 1$\sigma$ rms noise level in our final images is $\sim2\ \units{\mu Jy\ beam^{-1}}$.  We list the radio map noise level in Table \ref{tab:vladata}.

\subsection{Radio flux density measurements}
We required the source peak flux densities to be $>5\sigma_{\rm rms}$ in order to be considered detections.  Of the 8 sources we observed, we detected 5 of them.

Source fluxes and shapes were measured using the IMFIT task in CASA.  IMFIT models the emission of a compact radio source as a two-dimensional elliptical Gaussian.  We note that one of our sources, SDSS J095418.15+471725.1, has a complex radio morphology requiring a two-component Gaussian model in IMFIT. The second (dimmer) Gaussian component is located 0\farcs5 to the East ($\mathrm{PA} = 83^{\circ}$) and is unresolved within the $0\farcs43 \times 0\farcs31$ beam.  Futher deep, high-resolution multi-band radio imaging is required to fully ascertain the nature of the extended emission.  The luminosity and morphology of the radio emission in the available VLA data, however, are consistent with an extended jet arising from the central AGN.
This source is among the lowest-mass AGN known to have an extended radio jet \citep{nyland+17, 2019MNRAS.488..685M, 2020MNRAS.495L..71Y}.

\begin{figure*}
	\includegraphics[width=0.45\textwidth]{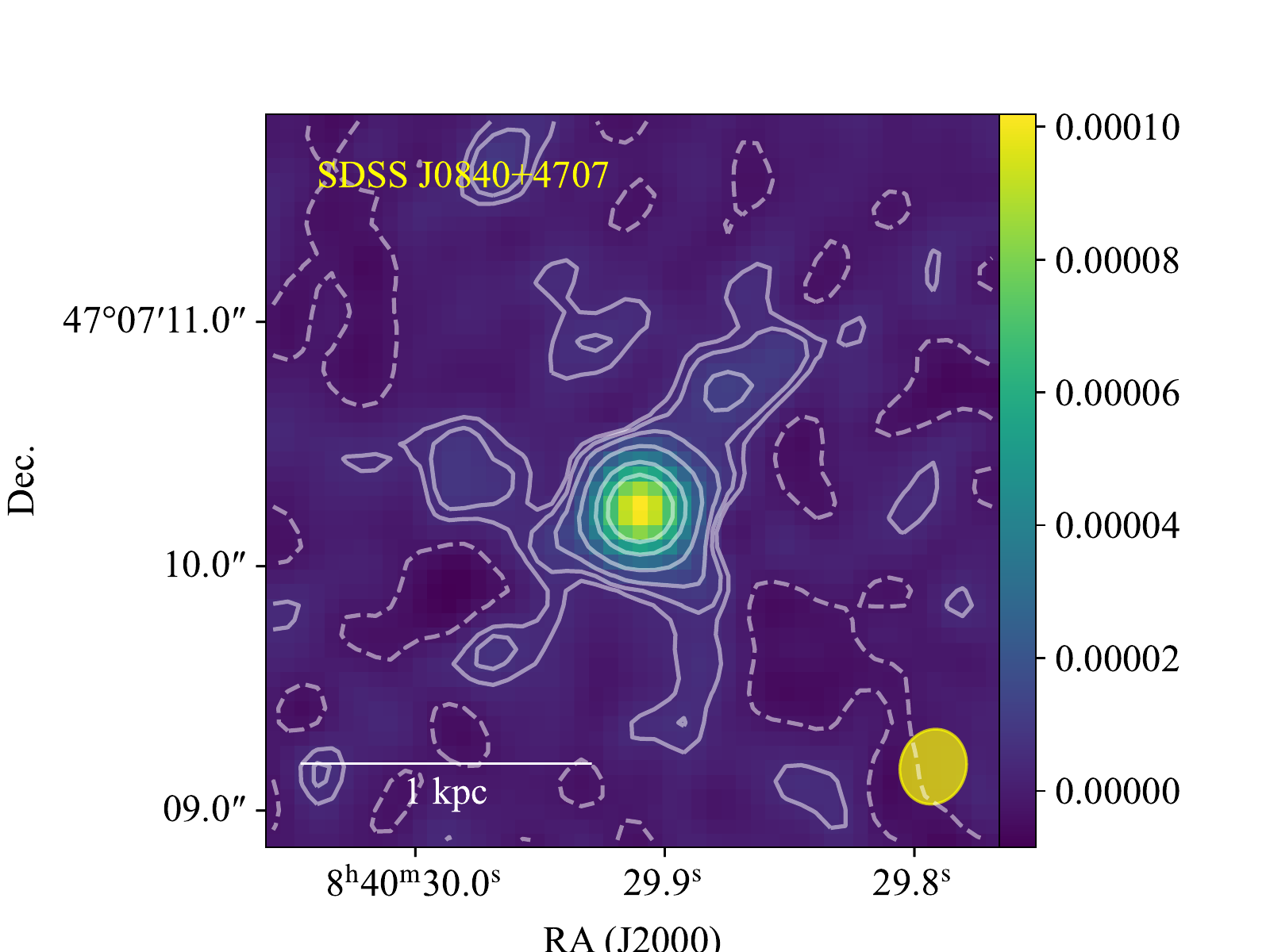}
	\includegraphics[width=0.45\textwidth]{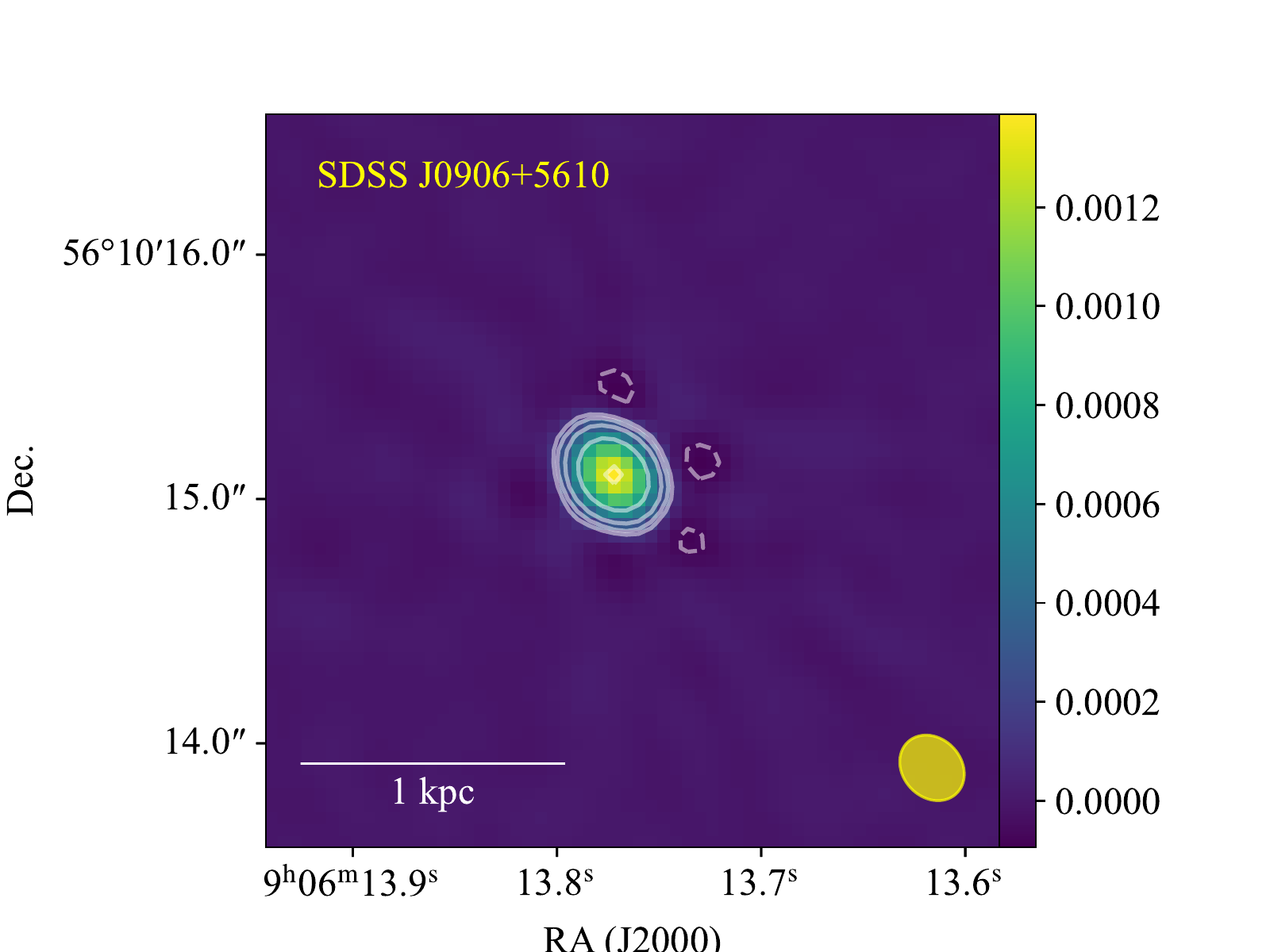}\\
	\includegraphics[width=0.45\textwidth]{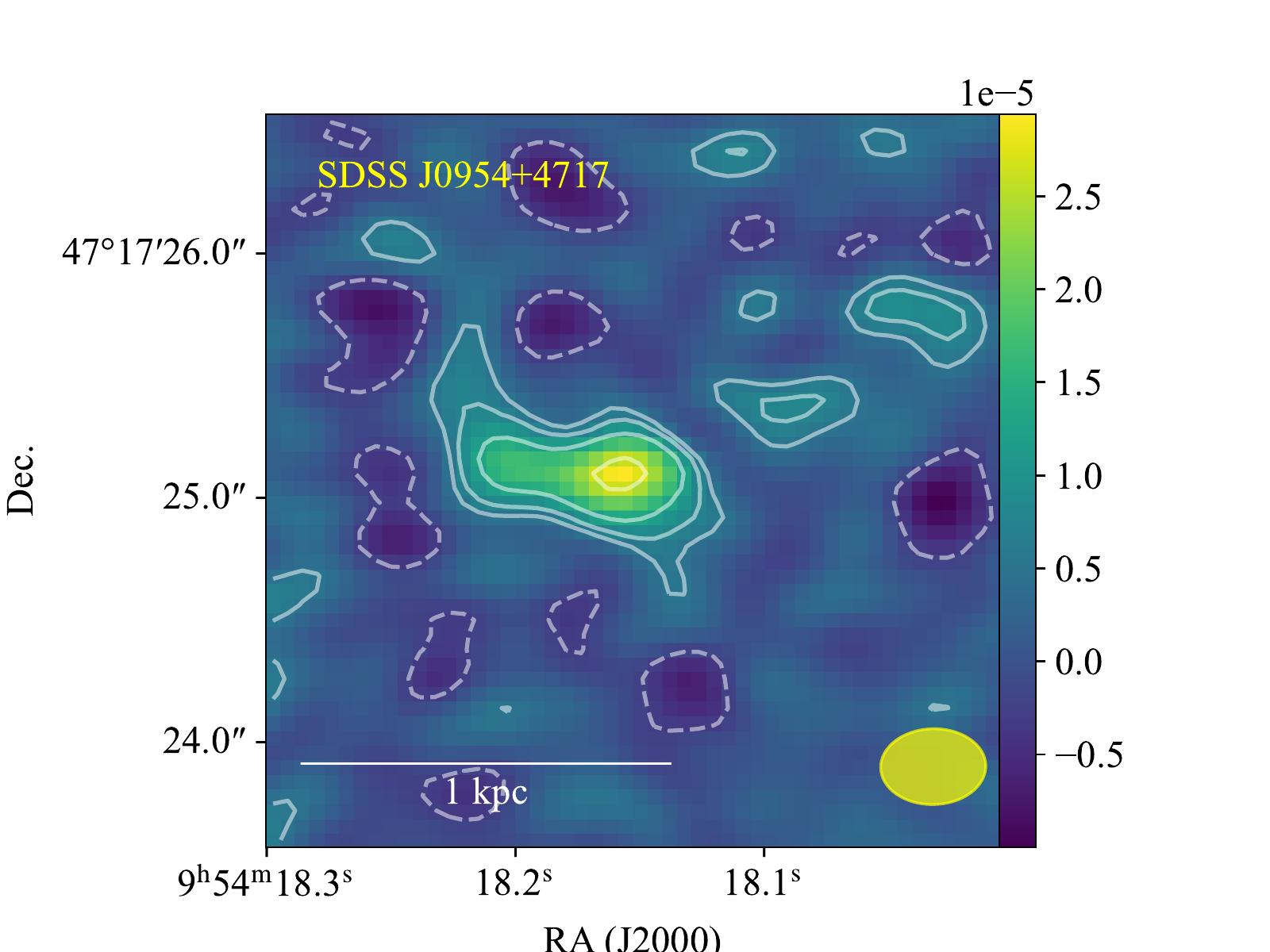}
	\includegraphics[width=0.45\textwidth]{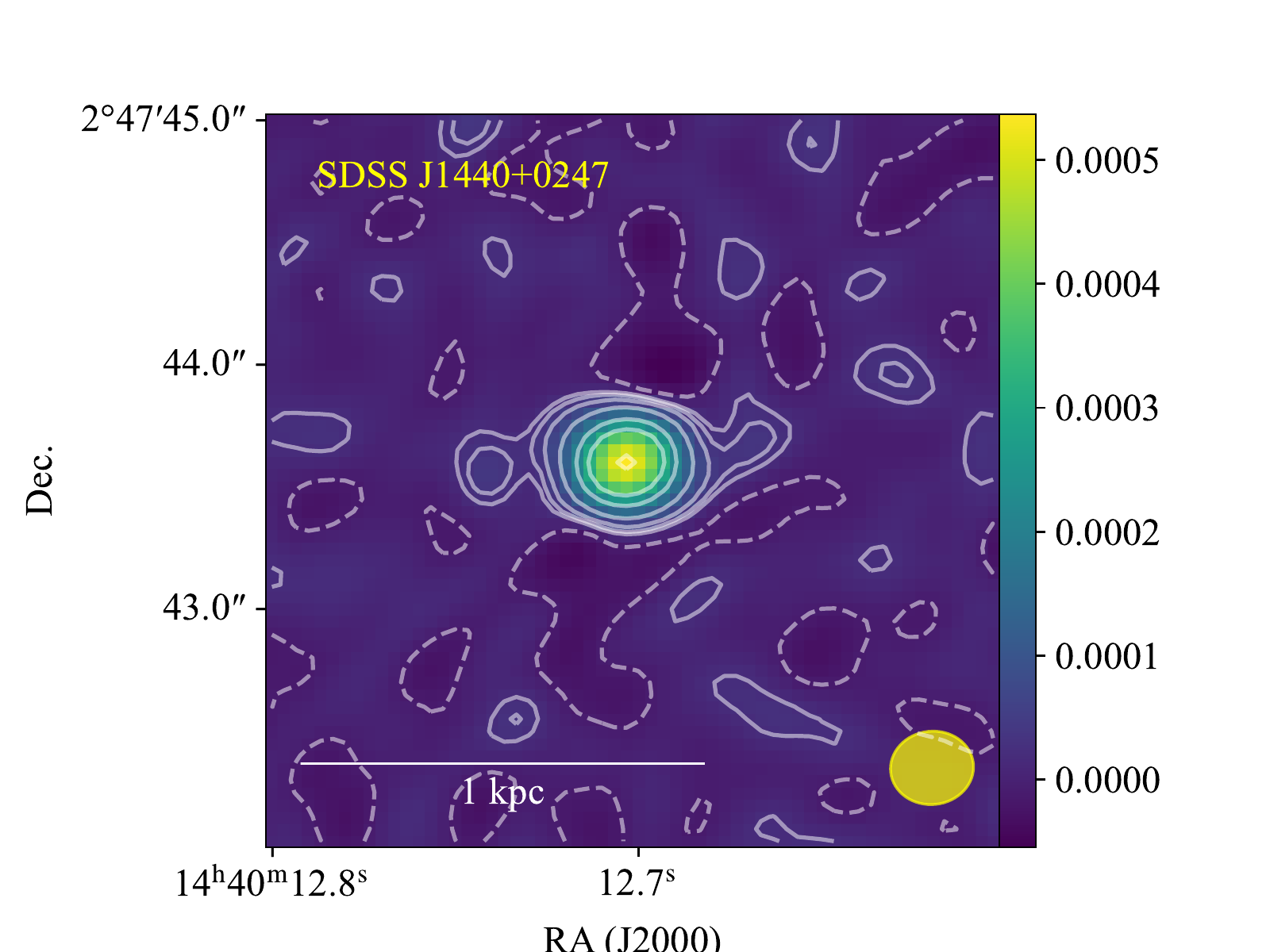}\\
	\includegraphics[width=0.45\textwidth]{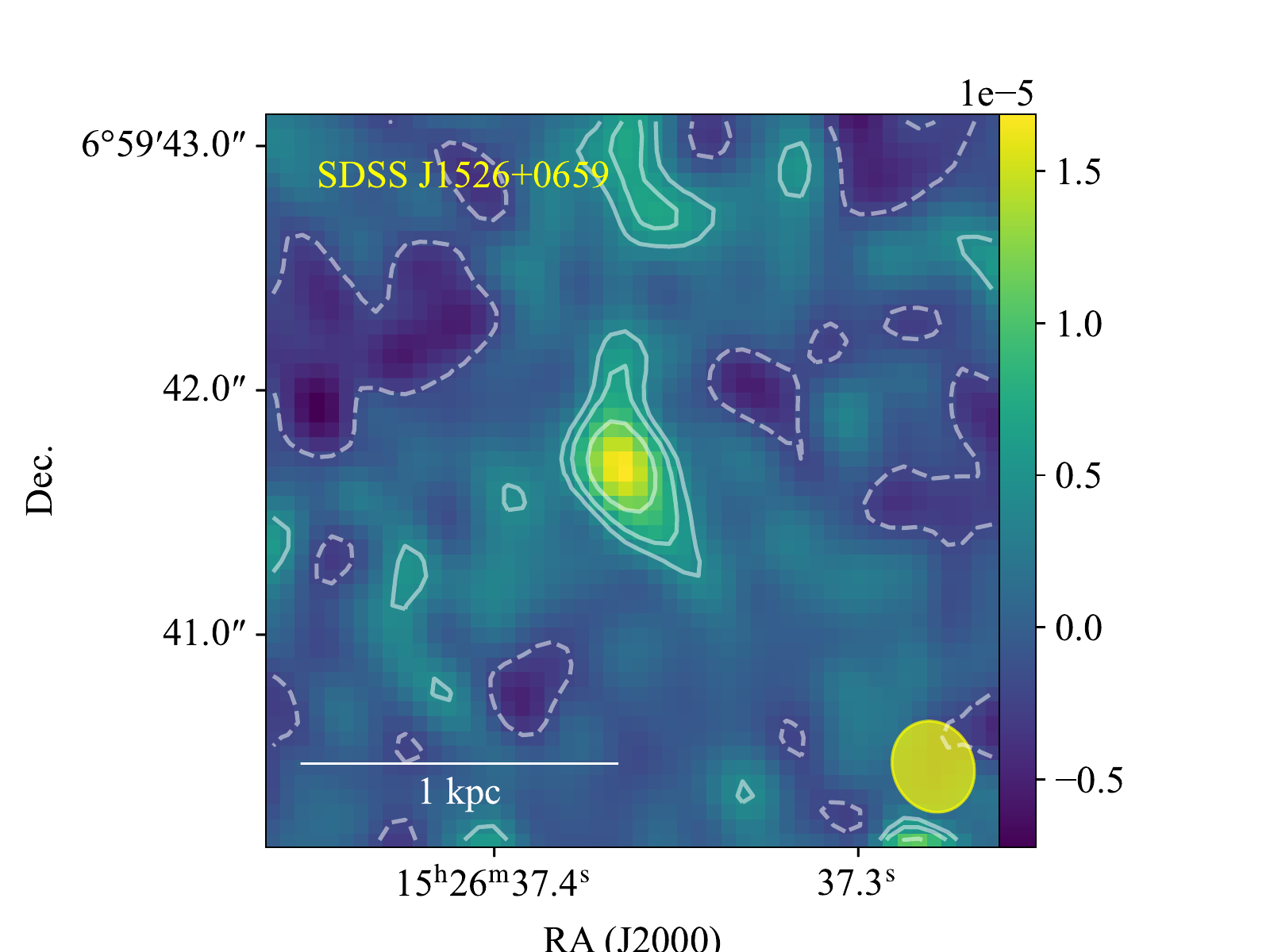}
    \caption{Radio maps of all 5 detected sources from our program.  For each panel, the name of each source is indicated in the top left; the synthesized beam is shown in the bottom right, and a scale bar indicating 1 kpc at the distance of the source is shown in the bottom left.  The color scale is different for each panel and is indicated by the colorbar at the right of each panel given in units of Jy.  Contours are drawn for panel at levels of ($-1$ [dashed], 2, 3, 5, 10, 20, 30, 50, and 80) times the map rms.  All sources plotted are clearly detected at high significance and are consistent with a single point source except for SDSS J0954+4717, which shows clearly extended emission.}
    \label{fig:vlaimages}
\end{figure*}

All other sources were unresolved or only marginally resolved by IMFIT, and are thus consistent with compact radio morphologies.  The flux density errors include a standard 3\% uncertainty in the absolute flux scale \citep{2017ApJS..230....7P}  added in quadrature with the errors reported by IMFIT.  A summary of the source properties is provided in Table~\ref{tab:vladata} and image cut-outs of the detected sources are shown in Figure~\ref{fig:vlaimages}.

% Main Imaging figure

% Main radio table
\begin{table*}
	\centering
	\caption{VLA source list and imaging results. In addition to the core radio flux noted below, SDSSJ095418.15+471725.1 has extended emission with a second, dimmer unresolved component with peak flux density $14.5 \pm 2.2\ \units{\mu{Jy}\ beam^{-1}}$ and integrated flux density $12.5 \pm 3.5\ \units{\mu{Jy}}$.}
	\label{tab:vladata}
	\begin{tabular}{lrrrrrr} 
		\hline
		%Source name & $D_L$ & Peak $S_{\nu}$          & Integrated $S_{\nu}$ & Map rms & Clean beam size & Clean beam PA\\
		Source name & $D_L$ & Peak $S_{\nu}$          & Integrated $S_{\nu}$ & $\sigma_{\rm rms}$ & Clean beam size & Clean beam PA\\
		            & (Mpc) & (\units{{\mu}Jy\ beam^{-1}}) &(\units{{\mu}Jy}) & (\units{{\mu}Jy}) & $\arcsec\times\arcsec$ & $^{\circ}$ E of N\\
		\hline
		SDSS J024656.39$-$003304.8 & 206 & \dots           & \dots        & 2.4 & \dots              & \dots \\ % RGG 1 AGN
		SDSS J084029.91+470710.4   & 188 & $101 \pm 2$     & $142 \pm 5$  & 2.2 & $0.31 \times 0.27$ & $-16$ \\ % RGG B Star forming
		SDSS J085125.81+393541.7   & 183 & \dots           & \dots        & 2.1 & \dots              & \dots \\ % RGG 48 Composite
		SDSS J090613.76+561015.1   & 209 & $1451 \pm 5$    & $1444 \pm 9$ & 65  & $0.29 \times 0.24$ & 43    \\ % RGG 9 AGN
		SDSS J095418.15+471725.1   & 145 & $27 \pm 2$      & $35 \pm 5$   & 3.0 & $0.43 \times 0.31$ & $-89$ \\ % RGG 11 AGN
		SDSS J144012.70+024743.5   & 132 & $546 \pm 10$    & $540 \pm 18$ & 1.0 & $0.34 \times 0.30$ & $-84$ \\ % RGG 32 AGN
		SDSS J152637.36+065941.6   & 171 & $15 \pm 2$      & $23 \pm 5$   & 2.1 & $0.38 \times 0.33$ & 23    \\ % RGG 119 Composite
		SDSS J160531.84+174826.1   & 140 & \dots           & \dots        & 2.3 & \dots              & \dots \\ % RGG 127 Composite
		\hline
	\end{tabular}
\end{table*}

\section{Results and Analysis}
\label{discuss}
For comparison with the fundamental plane, we combine our new radio data with X-ray data from \citet{2017ApJ...836...20B}  and black hole mass estimates from \citet{2013ApJ...775..116R}.  For a uniform analysis, we convert all values to our adopted luminosity distances listed in Table \ref{tab:vladata}.  The conversion is much smaller than the uncertainities in all of the quantities. We also convert the 90\% uncertainties in X-ray flux measured by \citet{2017ApJ...836...20B} to 68\% uncertainties.  Table 2 lists the logarithmic black hole mass ($\log{(M/\msun)}$), logarithmic 5 GHz radio luminosity ($\log{L_R}$), and logarithmic 2--10 keV X-ray luminosity ($L_X$) for each of our sources.  Note that $L_R = \nu L_{\nu}$ and $L_X$ is a bandpass luminosity.  We adopt an uncertainty of 0.4 dex in mass as argued by \citet{2017ApJ...836...20B}.

\begin{figure*}
	\includegraphics[width=0.49\textwidth]{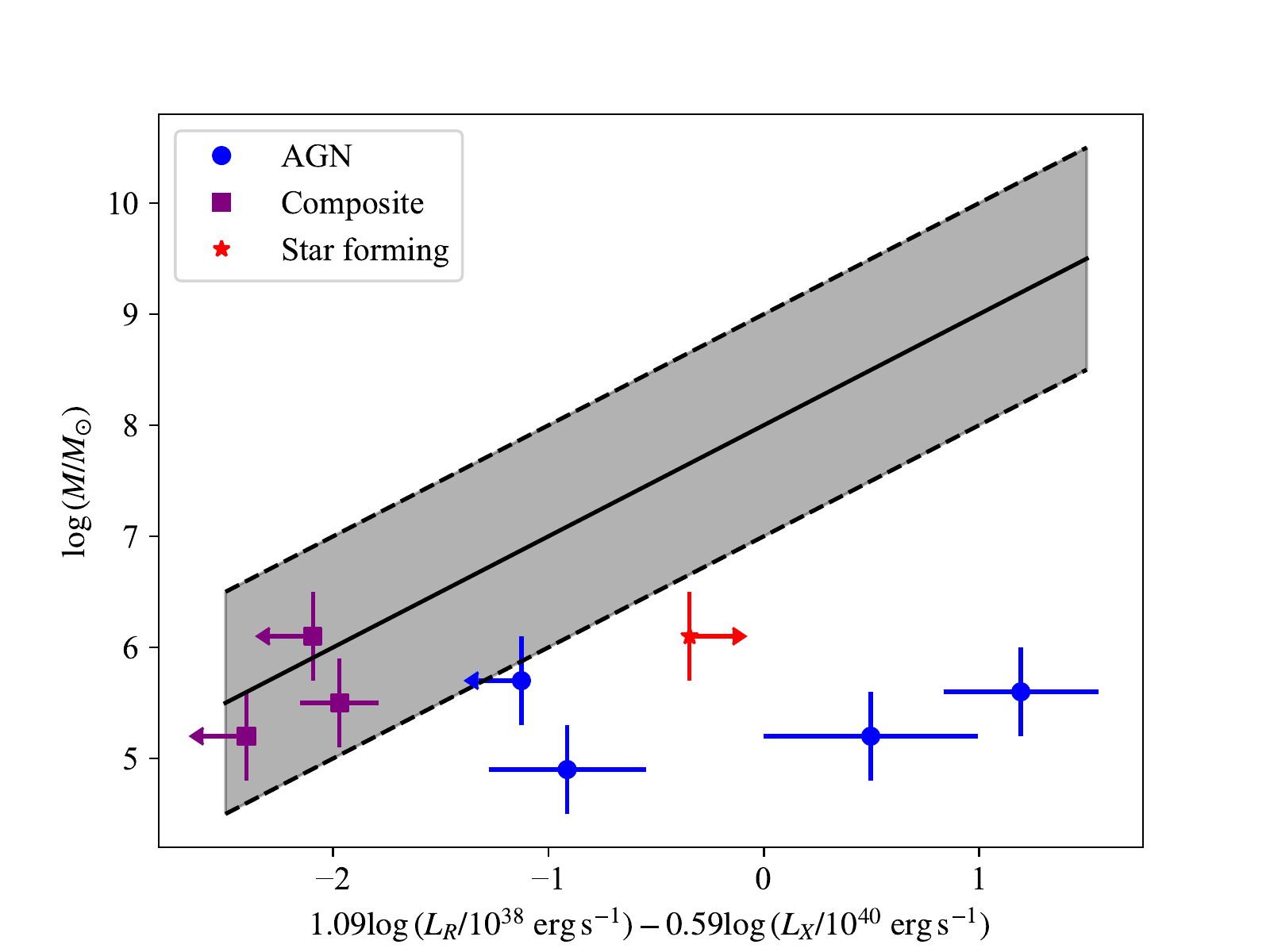}
	\includegraphics[width=0.49\textwidth]{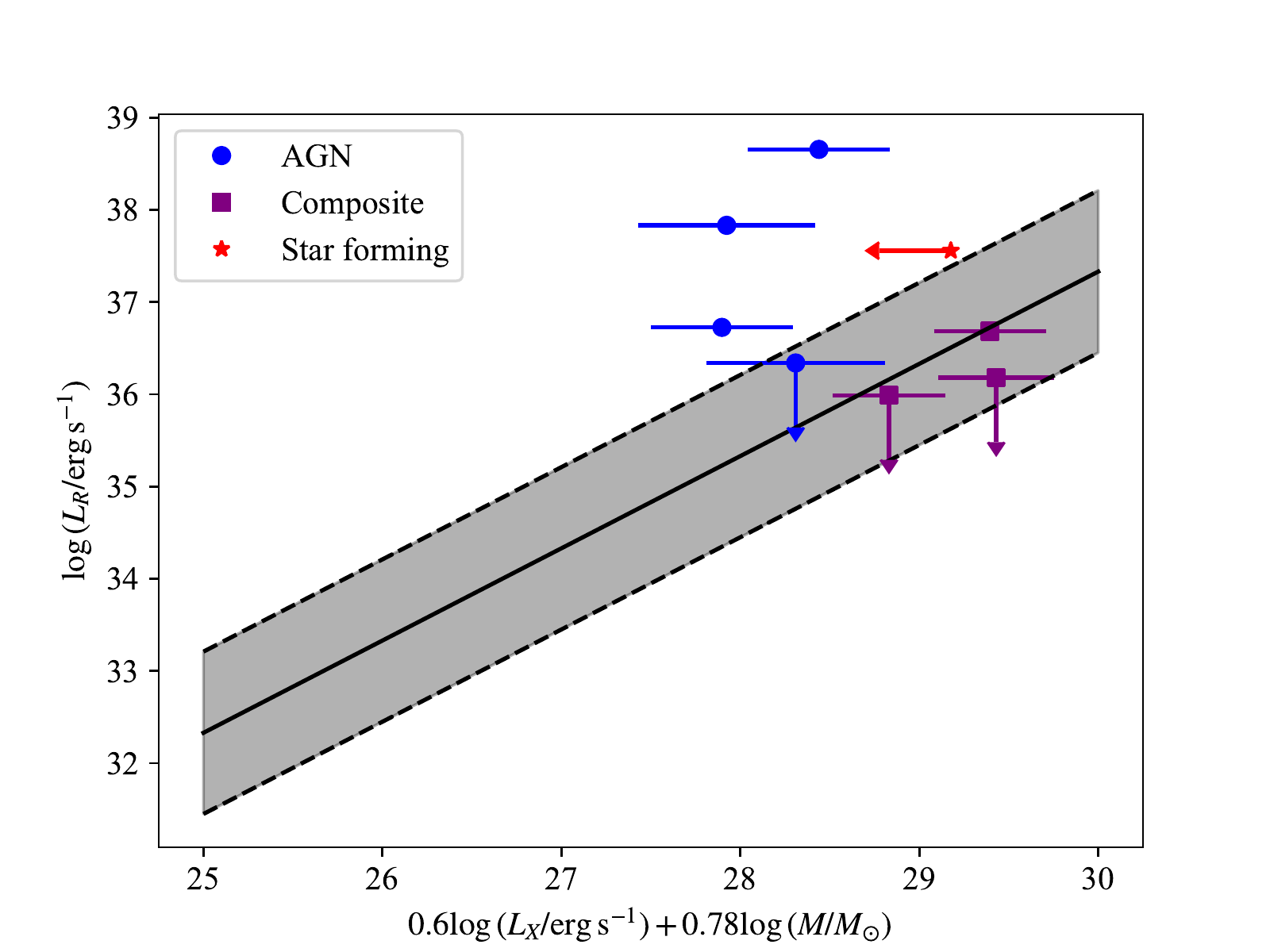}\\
    \includegraphics[width=0.49\textwidth]{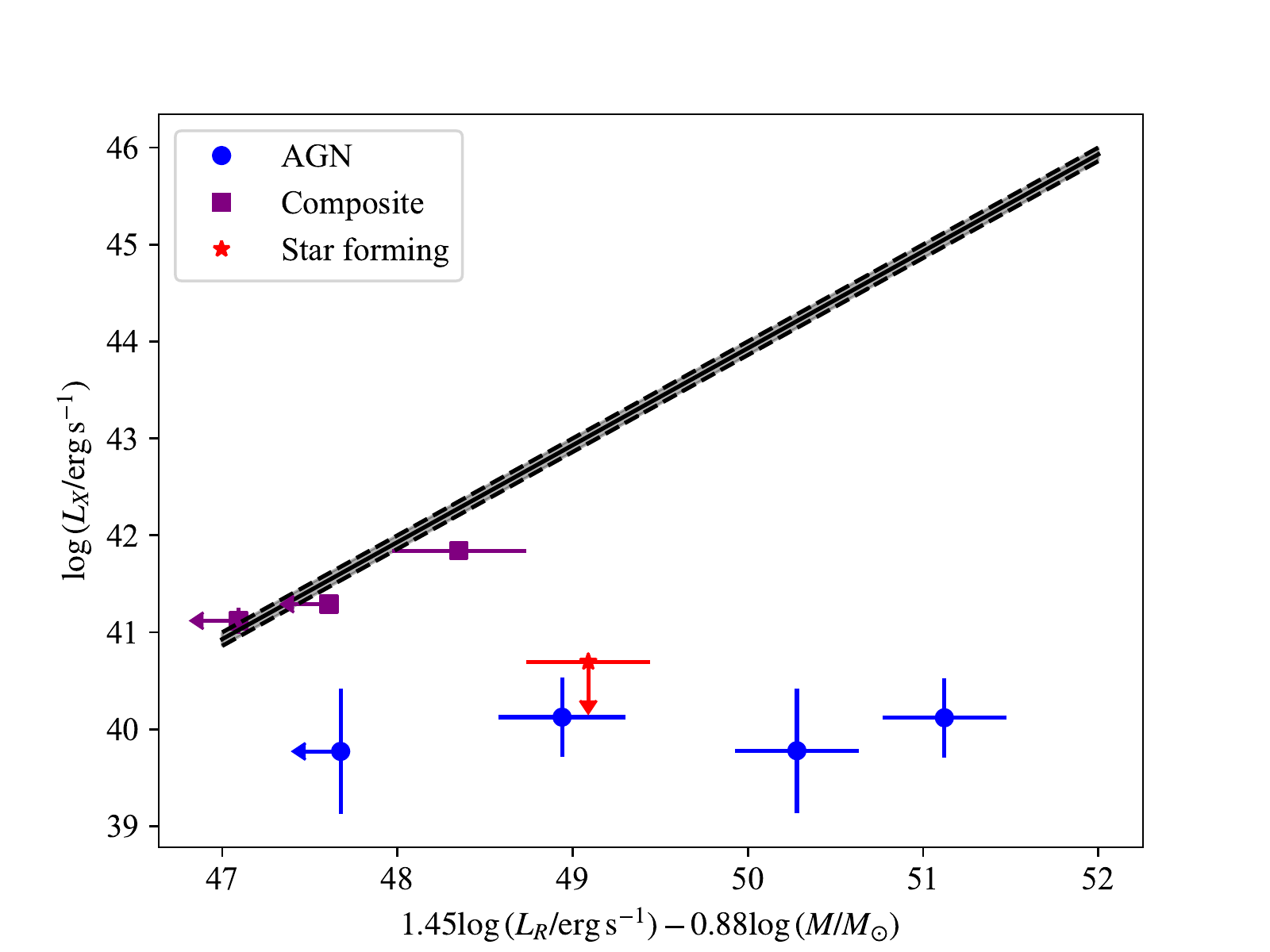}
	\includegraphics[width=0.49\textwidth]{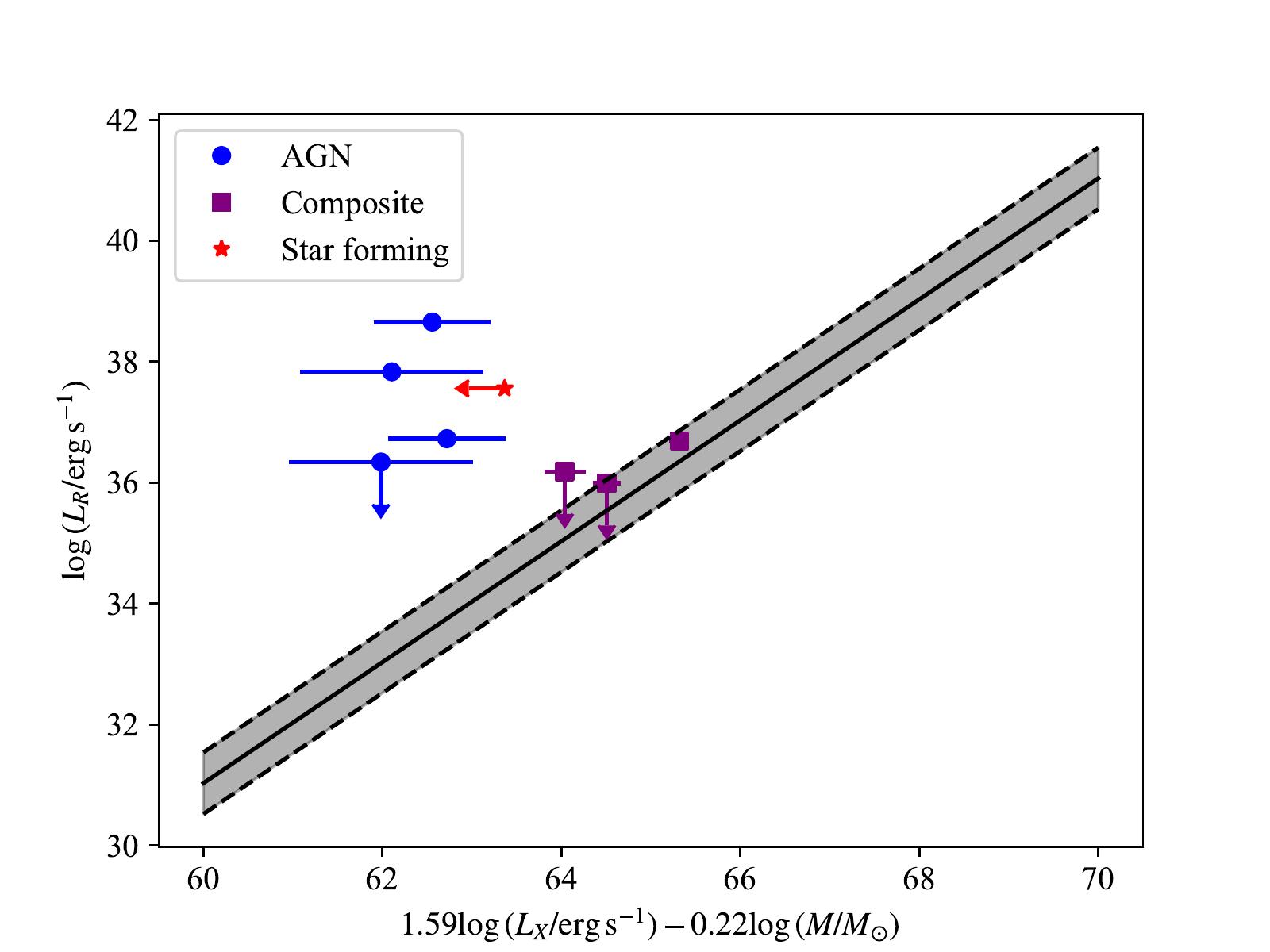}
\caption{\kgbf{IMBHs from our sample compared to several different fits of the fundamental plane of black hole accretion.  Each source is plotted with a symbol and color based on the BPT line diagnostic classification due to \citet{2013ApJ...775..116R}. Mass data come from \citet{2013ApJ...775..116R}, and X-ray data come from \citet{2017ApJ...836...20B} with $1\sigma$ uncertainties shown.  Radio upper limits of the AGN and composite sources are shown with an arrow at the $3\sigma$ limit.  The X-ray upper limit of the star-forming source is also shown at the $3\sigma$ value.  The directions of the arrows for limits is determined by the choice of independent variable and sign of the slopes of the dependent variables.  In each panel, the black line shows the best fit of the given work, and the shaded area bounded by black dashed lines is the $1\sigma$ intrinsic scatter in the direction of the ordinate. 
We do not invert the plane to a common ordinate because the fitted intrinsic scatters only have meaning in the directions given. Clockwise from top-left, the fits are due to  \citealt{2019ApJ...871...80G}, \citealt{2003MNRAS.345.1057M}, \citet{2012MNRAS.419..267P}, and
\citet{2014ApJ...787L..20D}.
Regardless of which result is used, the composite sources are consistent with the fundamental plane.  The star-forming source and three of the four AGN sources are inconsistent with the fundamental plane.}}
    \label{fig:fp}
\end{figure*}

Compared to the expected trend from higher-mass SMBHs and stellar-mass black holes, the IMBH are not in broad agreement. 
We plot the values for the IMBHs against the edge-on view of the best-fit fundamental plane from \citet{2019ApJ...871...80G} in Fig.\ \ref{fig:fp} along with the best fit and 1 dex intrinsic scatter.    While the black hole masses only span 1.2 dex, the abscissa of the edge-on projection spans over 4 dex.  Only four of the data points (including all 3 radio upper limits) are consistent with the fundamental plane and $1\sigma$ scatter within the measurement uncertainties.   We note that the one galaxy with emission lines classified as coming from star formation is also an X-ray upper limit and inconsistent with the best-fit fundamental plane.  

\kgbf{Comparing the data to other fits to the fundamental plane show similar offsets as seen in Fig.\ \ref{fig:fp}.  For example, \citet{2012MNRAS.419..267P} used a sample of sub-Eddington black holes (X-ray binaries and AGNs) with flat or inverted radio spectra, including de-beamed BL Lac objects but excluding Fanaroff-Riley type I galaxies. This sample results in a very tight relation with an intrinsic scatter of only 0.07 in the $L_{X}$ direction.  Our data similarly are discrepant for the AGN sources and the star-forming source while plausibly consistent for the composite sources. \citet{2014ApJ...787L..20D} used a sample of radiatively efficient black holes  (X-ray binaries and AGNs) to fit a fundamental plane relation with an intrinsic scatter of 0.51 in the $L_{R}$ direction.  \citet{2003MNRAS.345.1057M} used a sample of X-ray binaries, Seyferts, and quasars.  In all cases, the composite sources are consistent but the AGN sources are off the best-fit relation by several orders of magnitude.}

To better understand why the IMBHs are, overall, in disagreement with the fundamental plane, we consider whether any one of the measurements of $M$, $L_X$, or $L_R$ could be deficient for our purposes.  We consider each of these in turn but ultimately find that the measurements are sound.

\subsection{Reliable black hole mass measurements}
\label{reliablemass}
We find the black hole mass estimates not to be the reason for discrepancy between our data and the fundamental plane even though black hole mass estimation with single-epoch broad-line measurements are always subject to relatively large systematic uncertainties.  Single-epoch measurements use the width of a broad emission line as a measure of the velocity and the measurement of the continuum luminosity and the $R$--$L$ relation to estimate the size of the broad line \citep{2014SSRv..183..253P}.  The $R$--$L$ relation comes from reverberation mapping mass estimates of many AGN and shows that there is an empirical correlation between the size of the broad line region and the continuum luminosity \citep{2006ApJ...644..133B, 2009ApJ...697..160B, 2013ApJ...767..149B}.  The reverberation mapping masses are mostly calibrated to the $M$--$\sigma$ relation of quiescent galaxies with dynamical estimates of the black holes \citep[e.g.][]{2009ApJ...698..198G}.  This calibration is necessary because most reverberation mapping masses do not have information about the details of the kinematics and geometry (including viewing angle) of the broad line region.  \kgbf{Modeling of the broad line region kinematics and geometry is, however, possible and efforts to do so with as many AGN as possible is ongoing \citep[e.g.,][]{2011ApJ...730..139P, 2014MNRAS.445.3055P, 2022arXiv220603513B}.} There is additional potential for systematic uncertainty in that many of the host galaxies of the reverberation mapping AGN are pseudobulges, which do not follow the low-scatter $M$--$\sigma$ relation that classical bulges and elliptical galaxies do \citep{2004ARA&A..42..603K, 2008ApJ...688..159G, 2010ApJ...716..942F, 2011ApJ...733L..47F, 2013ARA&A..51..511K, 2020ApJ...889...14Y, 2020MNRAS.493.1686L, 2020ApJ...897..102C}.  Finally, because only H$\alpha$ broad lines are measured in our sources, an additional calibration has to be done from the H$\alpha$ $R$--$L$ relation to the H$\beta$ $R$--$L$ relation \citep{2013ApJ...775..116R}.  

Despite the fact that there is the potential for large systematic uncertainty in the mass estimates, we discount this as the primary\,---\,or even a significant\,---\,reason for the discrepancy between the IMBH measurements and the fundamental plane.  We discount this possibility because it would require two sources to have black hole masses of $\sim 3 \times 10^{7}$--$10^{10}\ \msun$, despite the fact that the host galaxy stellar masses are $M_{*} < 3 \times 10^{10}\ \msun$.  

\subsection{Reliable X-ray luminosity measurements}
\label{reliablexray}
The possibility of incorrect X-ray luminosity measurements is similarly implausible.  Because the data in Fig.\ \ref{fig:fp} are systematically to the right of the best-fit relation and because $\xi_{M,X} = -0.59 < 0$, to account for the discrepancy the X-ray luminosities would have to be underestimated.  There are two possibilities in which this could happen: inaccurate measurement of X-ray flux or accurate measurement of X-ray flux but incomplete accounting of absorption.  In either case the magnitude of the discrepancy makes this implausible.  The discrepant points are \kgbf{off by} $\sim{1}$--$3$ dex and thus the X-ray luminosity would have to be underestimated by $1.5$--$5$ dex.  That amount of absorption is not seen in the X-ray spectral analysis.   SDSS J095418.15+471725.1 had enough X-ray counts to model the total amount of absorption and was found to be a column of $N_{\mathrm{H}} \approx 10^{21}\ \units{cm}^{2}$ \citep{2017ApJ...836...20B}, which leads to a factor of a few correction to the unabsorbed hard X-ray flux (which we use in our analysis).  SDSS J144012.70+024743.5 and SDSS J090613.76+561015.1 have similar hardness ratios ($(H - S) / (H + S) \approx -0.7$) and are inconsistent with typical Compton-thick AGN spectrum. 
Thus, overall we find no reason for a significant defect in the X-ray luminosity measurements.

\subsection{Reliable radio luminosity measurements}
\label{reliableradio}
\kgbf{We will also argue that} our radio luminosity measurements are robust measurements of emission from the central AGN of our sample.  An important thing to consider for radio observations of sources like our sample is contamination in the radio data.  Since the discrepant points are to the right of the best-fit relation, if non-AGN radio emission were included in the radio flux measurements, it would cause them to move to the right in Fig.\ \ref{fig:fp}.  The most likely potential source of contamination is from star formation, but we can rule this out based on (i) our experimental setup, (ii) star-formation rates and AGN activity based on optical line luminosities, and (iii) the BPT classification of our sources.  

\subsubsection{Experimental setup}
Our experimental setup was designed to exclude star formation to the extent possible by using the most extended array possible with VLA at a relatively high frequency (C band, 4--8 GHz) for the highest angular resolution possible with VLA ($\sim0\farcs3$).  All core radio emission is unresolved (except for the extension in SDSS J0954+4717).  At the distances of the sources, the core radio emission is thus confined to 200--300 pc and therefore morphologically consistent with only AGN emission and inconsistent with extended star formation.  

\subsubsection{Optical-line estimates of star-formation rates and AGN activity}
Our estimates of continuum radio emission resulting from star-formation-linked emission lines indicates that the radio emission observed either mostly or entirely originates from AGN activity.
\citet{2017ApJ...836...20B} estimate conservative upper limits to the star formation rates by assuming that all of the narrow-line H$\alpha$ emission comes from star formation.  These upper limits are sufficient to show that the X-ray flux is still dominated by AGN emission (with the notable exception of SDSS J084029.91+470710.4, the one star-forming source in the sample).  These upper limits to the star formation, however, are too conservative to rule out radio emission as arising from star formation as they predict a high amount of star formation.
A more detailed analysis of the relative contribution by AGN and star-formation to the H$\alpha$ narrow lines would be necessary to estimate the star-formation contamination in AGN. 
Another indication that our radio measurements are not contaminated by star formation  comes from the relation between [\ion{O}{III}] luminosity and AGN radio luminosity.  This relation connects AGN narrow emission line luminosity with radio power.  There is a considerable scatter in such measured relations \citep[e.g.,][]{2016A&A...591A..88B}, but based on the [\ion{O}{iii}] luminosities measured by \citet{2013ApJ...775..116R}, the predicted range of AGN radio luminosities is at or above what we observe, even for radio quiet AGN.  That is, there is no need to invoke star formation to explain the radio, based on the [\ion{O}{iii}] flux.  For small galaxies like those in our sample, typical star-formation-induced radio luminosities are much lower than $10^{37}\ \units{erg\ s^{-1}}$ \citep{2020ApJ...888...36R}.

\subsubsection{BPT classification}
Finally, the fact that the most discrepant sources are AGN by BPT diagnostics suggests that star formation is only a minor contributor to the total emission.  The BPT diagnostics show that the contribution by star formation to the optical emission lines is subdominant to the contribution from AGN.  It is then reasonable to assume that the contribution to radio continuum emission is similarly dominated by AGN, but the timescales of AGN activity and star formation in these low-mass systems may complicate any inferences here.

Thus based on the totality of the above arguments in this subsection, we conclude that star formation contamination of the radio emission cannot be a large part of the discrepancy.  Nor can a combination of effects of the different variables result in the discrepancy of the AGN source, which systematically fall to the same side of the plane.  Of the three AGN sources with radio detections, even if half of the radio emission came from star formation and the masses were underestimated by 0.5 dex, they would be inconsistent with the fundamental plane. They would lie more than the estimated measurement uncertainty below the 1 dex intrinsic scatter of the relation.  

\section{Discussion}

The natural conclusion of our analysis is that the IMBHs with AGN-like line ratios do not, in general, follow the fundamental plane of black hole accretion.  It is unexpected to find that these particular sources do not follow the same relation when other, similar, sources do \citep{2014ApJ...788L..22G}.  As we have argued above in \S\S\ \ref{reliablemass}--\ref{reliableradio}, we have robust measurements of $M$, $L_X$, and $L_R$.  We now consider potential reasons that our sample would not follow the fundamental plane  despite having accurate data.

\subsection{Potentially optically thin radio emission}
One potential explanation is that the radio emission we detect \kgbf{is not optically thick emission expected to arise in the core of the jet but is instead optically thin radio emission from jet lobes.
Multiple interpretations of the fundamental plane rely on the observed radio emission to only be optically thick jet core emission, which would result in a flat or inverted radio spectrum \citep{2003MNRAS.343L..59H, 2004MNRAS.355.1105F, 2004A&A...414..895F, 2007MNRAS.381..589M, 2012MNRAS.419..267P}.  
The presence of optically thin lobe emission, which would result in a steep spectrum radio source, is possible if the lobes are small enough ($\lesssim 200\ \units{pc}$) to fit within the synthesized beam of our observations.  Alternatively, the radio emission we observe could be the product of an earlier period of jet activity compared to the observed X-ray emission.}  If the radio emission is actually relic emission from earlier activity, the radio emission currently observed comes from an aged population of electrons and will have a flatter spectral index.  

With our single band of radio coverage, we are not able to measure the spectral index with sufficient precision \kgbf{to determine if the radio emission is optically thick or thin.}  Observations at a higher and/or lower frequency band would be necessary.  If the only radio emission seen is from an earlier epoch of activity, then the current epoch of radio emission must be substantially lower and would bring the discrepant AGN points closer to the relation.

\kgbf{A likely explanation for seeing current X-ray activity but no or substantially lower core radio emission from the current epoch is that the jets have been quenched when the accretion is likely radiatively efficient as is seen in X-ray binaries \citep{2003MNRAS.344...60G} and other low-mass AGN \citep{2006ApJ...636...56G, 2007ApJ...656...84G}.  A necessary but not sufficient condition for radiatively efficient emission is a high Eddington ratio.  The Eddington ratios of our sample cannot be fully determined without full SED modeling, but we can put a limit on it by calculating the X-ray Eddington ratio ($L_X / L_{\mathrm{Edd}}$).  Most of our sources have $L_X / L_{\mathrm{Edd}} \lesssim 10^{-3}$, but two of them (SDSS J152637.36+065941.6 and SDSS J160531.84+174826.1) have $L_X / L_{\mathrm{Edd}} \approx 10^{-2}$.  Typical X-ray bolometric corrections are 10--50 \citep{2004MNRAS.351..169M, 2007MNRAS.381.1235V}.
Jet quenching at high Eddington rates would not explain why the sources with line ratios in the composite region of the BPT diagram are consistent with the fundamental plane.  It is possible that the composite AGN in our sample with only limits on their radio emission are also far off the plane, but we would need substantially deeper radio measurements to do so.  For composite sources, it is difficult to separate out relative contributions without spatially resolved spectroscopy, multiwavelength constraints on variability and hard X-ray data \citep[e.g.,][]{2021ApJ...910....5M}.  Deeper high-resolution multi-band radio imaging with ngVLA would also allow for better inference \citep{nyland+18}.  }

It is, however, worth noting that our sample was selected without consideration as to whether or not the AGN had radio emission.  In our small sample, the IMBHs with line ratios in the AGN region of the BPT diagram are, at face value, more likely to have detectable radio emission; but the sample size is a factor of $\sim3$ too small to make any conclusive statements about even the sample statistics, let alone the population of IMBHs.  

Finally, we note that high-Eddington quasars are not always jet-quenched.  The reasons for why radio jets are present or absent is a complicated question that involves jet-launching physics (including black hole spin and, potentially, mass) as well as other factors that influence the duty cycle and lifetime of the radio-loud phase. 
The energy distribution of radiating particles in the jets of black holes accreting at low rates may be thermal or dominated by a thermal component, as is seen in X-ray binaries \citep{2013MNRAS.434.2696S} and Sgr A* \citep{2001A&A...379L..13M, 2017SSRv..207....5R}.
New radio surveys \citep[e.g., VLASS][]{2020ApJ...905...74N} and telescopes \citep[DSA-2000 and ngVLA][]{2019BAAS...51g.255H, 2015arXiv151006438C, 2015arXiv151006411C, 2016SPIE.9906E..27M} will enable new tests of how jets are triggered and launched under different conditions, as well as their typical lifetimes.

\subsection{Estimating black hole masses}
It is clear from our results that an uncritical application of the fundamental plane to radio and X-ray data to estimate black hole masses will fail some of the time.  It is necessary to understand when it is appropriate to use the fundamental plane.  The sources for which the fundamental plane applies (at least as well as it does: $\sim1$ dex scatter; \citealt{2019ApJ...871...80G}) appear to be those that are decidedly accreting at rates where radiatively inefficient accretion flows (RIAFs) are expected to operate or, at the most, very low-level Seyfert-like activity.  

\kgbf{The preference for low accretion rates is expected based on models of the fundamental plane as a result of sub-Eddington jet-dominated emission \citep{2004A&A...414..895F} and on other radiatively inefficient accretion flow (RIAF) models that apply for X-ray emission that is radiatively inefficient.  In the jet-dominated models, both the radio and X-ray emission originate in the jet.  The radio is synchrotron emission, and the X-ray is synchrotron self-Compton emission.  There is observational support for these models to be correct based on analysis of best-fit fundamental plane slopes using a sample of X-ray binaries, low-luminosity AGN, and BL Lac objects \citep{2012MNRAS.419..267P}.   In such a model it is straightforward to see why there is a correlation between the three quantities.  If low-luminosity Seyferts (such as NGC 4388, NGC 4477, NGC4501, NGC 4698, or NGC 4945; \citealt{2019ApJ...871...80G}) are still jet-dominated in X-ray emission, then they would naturally be described by such models.  
In RIAF models, the X-ray emission is radiatively inefficient and can include synchrotron, inverse Compton scattering of photons originating in the accretion flow, and synchrotron self-Compton.  By their very nature, RIAF models do not operate when the accreting source has a luminosity at a higher fraction of its Eddington luminosity.  The fundamental plane predicted by several RIAF models has been computed and also compared to observations of a wide variety of sources \citep[e.g.,][]{2003MNRAS.344...60G, 2003MNRAS.343L..59H, 2003MNRAS.345.1057M, 2012MNRAS.419..267P}.}

Seyferts accreting at high rates and quasars (possibly including the IMBHs with AGN line ratios in this work) are likely to produce X-ray emission predominantly through inverse Compton scattering of ultraviolet photons originating from a thin or slim accretion disk.  
\kgbf{Sources with observational properties that we normally associate with radiatively efficient thin disk accretion such as broad emission lines would not be expected to follow the fundamental plane if either sub-Eddington jet models or RIAF models  are the correct model.}
High-accretion rate AGN are, however, still likely to show some correlation between radio luminosity, X-ray luminosity, and black hole mass, even if the correlation is different \citep{2020MNRAS.496..245Z}.  Additionally, \citet{2021ApJ...906...88F} found that at Very Large Baseline Array (VLBA) resolution radio emission associated with Seyfert sources  may be contaminated by jet-driven interactions with the interstellar medium.

\subsection{Need for an independent estimate of Eddington fraction}
In order to use the fundamental plane to estimate black hole mass, one therefore needs, at a minimum, an estimate of the Eddington fraction to determine roughly whether a given accreting black hole is accreting at $f_{\mathrm{Edd}} \lesssim 10^{-2}$.  Determining $f_{\mathrm{Edd}}$ from just an X-ray spectrum alone is not possible because the X-ray photon index--Eddington fraction ($\Gamma$--$f_{\mathrm{Edd}}$) relation is not monotonic \citep{2008ApJ...682...81S, 2009ApJ...690.1322W, 2012ApJ...749..129G, 2017MNRAS.470..800T}.  The relation has negative slope at $f_{\mathrm{Edd}} \lesssim 10^{-2}$ and positive slope at higher accretion rates and thus does not break the ambiguity.  

\subsubsection{\texorpdfstring{$\alpha_{\mathrm{OX}}$}{Alpha\_OX}}
One way to break the ambiguity would be by measuring $\alpha_{\mathrm{OX}}$, the slope of the AGN SED between 2500\ \AA\ and 2 keV.  The value of $\alpha_{\mathrm{OX}}$ correlates positively with Eddington fraction \citep{2010A&A...512A..34L} so that large values would indicate an AGN likely accreting as a thin disk and thus the fundamental plane should not be used to estimate mass. 

At low masses such as the sources in this study, $\alpha_{\mathrm{OX}}$ has been shown to be flatter than expected by extrapolations from higher masses \citep{2012ApJ...761...73D}.  A low value of $\alpha_{\mathrm{OX}}$, however, could indicate either low accretion rates\,---\,and thus an appropriate venue for fundamental plane mass estimation\,---\,or significant obscuration.  Significant obscuration could be inferred from X-ray spectral fitting.   \citet{2017ApJ...836...20B} explored these objects in terms of where they sit on the  $\alpha_{\mathrm{OX}}$--$f_{\mathrm{Edd}}$ relation as well as on the relation between $\alpha_{\mathrm{OX}}$ and the 2500\ \AA\ luminosity ($L_{2500}$). They found that $f_{\mathrm{Edd}}$ was higher for the composite objects than for the AGN objects. The composite objects also have higher values of $\alpha_{\mathrm{OX}}$.   Additionally and of most relevance, the composite objects (i.e., the objects consistent with the fundamental plane) are also consistent with the $\alpha_{\mathrm{OX}}$--$L_{2500}$ relation that is found for more massive AGN. The objects that are inconsistent with the $\alpha_{\mathrm{OX}}$--$L_{2500}$ relation are also inconsistent with the fundamental plane.  \kgbf{The deviations from the $\alpha_{\mathrm{OX}}$--$L_{2500}$ may be a result of intrinsically X-ray weak AGN at high-Eddington rates as has been seen in a number of higher-mass sources \citep[e.g.,][]{2007ApJ...663..103L, 2015ApJ...805..122L}.}

\subsubsection{SED modelling}
A related way to estimate Eddington fraction is full spectral energy density (SED) modeling.  Compared to single-band methods, SED modeling is relatively resource intensive in that it requires as many bands of obsservations as possible, but it can be used to infer Eddington fraction \citep[e.g.,][]{1999ApJ...516..672H, 2010ApJS..187..135E, 2011ApJ...735..107M, 2012A&A...539A..48M, 2012MNRAS.422..478R, 2014MNRAS.438.2804N}.

\subsubsection{X-ray variability}
Another potential possibility is to use X-ray variability, which correlates with mass and $f_{\mathrm{Edd}}$ \citep{2003MNRAS.345.1271V, 2006Natur.444..730M, 2020MNRAS.492.1363P} along with the photon index of the X-ray spectrum, which correlates with $f_{\mathrm{Edd}}$ \citep{2008ApJ...682...81S, 2009ApJ...690.1322W, 
2012ApJ...749..129G, 
2017MNRAS.470..800T}.  Both of these relations have substantial intrinsic scatter such that they may only be useful as a rough guide.

\begin{table}
	\centering
	\caption{Fundamental plane data for analysis.  Masses are from \citet{2013ApJ...775..116R}.  Radio data are $L_R \equiv \nu L_{\nu}$ at $\nu = 5 \units{GHz}$ from this work.  X-ray data are 2--10 keV bandpass luminosities from \citet{2017ApJ...836...20B}.}
	\label{tab:fpdata}
	\begin{tabular}{lrrr} % four columns, alignment for each
		\hline
		Source name & $\log{M}$ & $\log{L_R}$ & $\log{L_X}$\\
		            & \msun & \units{erg\ s^{-1}} & \units{erg\ s^{-1}}\\
		\hline
		SDSS J024656.39$-$003304.8 & 5.7 & $<36.34$          & $39.8  \pm 0.6$  \\ % RGG 1 AGN
		SDSS J084029.91+470710.4   & 6.1 & $37.56 \pm 0.02$  & $<40.7$          \\ % RGG B Star forming
		SDSS J085125.81+393541.7   & 6.1 & $<36.18$          & $41.1  \pm 0.1$  \\ % RGG 48 Comosite
		SDSS J090613.76+561015.1   & 5.6 & $38.66 \pm 0.02$  & $40.1  \pm 0.4$  \\ % RGG 9 AGN
		SDSS J095418.15+471725.1   & 4.9 & $36.73 \pm 0.06$  & $40.1  \pm 0.4$  \\ % RGG 11 AGN
		SDSS J144012.70+024743.5   & 5.2 & $37.83 \pm 0.01$  & $39.8  \pm 0.6$  \\ % RGG 32 AGN
		SDSS J152637.36+065941.6   & 5.5 & $36.68 \pm 0.10$  & $41.84 \pm 0.04$ \\ % RGG 119 Composite
		SDSS J160531.84+174826.1   & 5.2 & $<35.99$          & $41.29 \pm 0.08$ \\ % RGG 127 Composite
		\hline
	\end{tabular}
\end{table}

\section{Summary}
\label{concl}

In this work, we examined a sample of 8 AGN identified to have low-mass black holes ($\log{M} \le 6.1$) based on H$\alpha$ single-epoch estimates.  Seven of these had been identified as AGN and one had been identified as a star-forming galaxy based on their optical narrow emission line ratios.  Each source was observed for the first time with VLA for 3--4.5 hours at $C$ band (4--8 GHz) in A configuration with typical angular resolution of $\sim0\farcs3$.  Of the 8 sources, 5 were unambiguously detected at a minimum S/N of $5\sigma$, and 3 were undetected with typical map rms of $2.3\ \units{\mu Jy}$.  Four of the detected sources were consistent with a point source, and one source (SDSS J0954+4717) has clear extension that can be modeled as two point sources while morphologically resembling a jet.  With a black hole mass estimate below $10^5\ \msun$, SDSS J0954+4717 is one of the smallest black holes to show evidence of an extended jet.

Combining our new radio observations with existing X-ray data and the mass estimates, we compare the sample with the fundamental plane of black hole accretion.  The lone star-forming source (SDSS J084029.91+470710.4), while detected in X-rays, is undetected in the 2--10 keV band used for the fundamental plane and is thus inconsistent with the fundamental plane, as would be expected for a star-formation-powered source.  All three sources with narrow line ratios indicating composite star-forming/AGN-powered narrow lines are consistent with the fundamental plane, though two of them are undetected in our radio observations.  One of the four sources with narrow line ratios indicated AGN-powered narrow lines is consistent with the fundamental plane, but it is also undetected in our radio observations.  The remaining three sources with AGN-powered narrow lines are inconsistent with the fundamental plane, despite its 1 dex scatter.  

All three inconsistent sources are off of the plane by a large amount with values of $L_R$ 1--3 orders of magnitude too large or with values of $L_X$ 2--6 orders of magnitude too small.
We argue that this inconsistency is not attributable to the misidentification of star-formation flux to AGN activity.  Nor can the discrepancy be rectified by appealing to larger black hole masses as it would require implausibly large black hole masses.  We thus conclude that the fundamental plane of black hole accretion does not apply to these sources.  \kgbf{To ascertain what, specifically, about these sources makes them unfit for the fundamental plane requires more observational investigation.  In particular, improvements could be made with multi-band radio observations to determine if the radio spectrum is steep or flat/inverted, deep VLBA observations to see if any of the emission is extended on scales smaller than 0\farcs3, or broadband SED modeling to infer the radiative efficiency of the IMBHs in our sample.   Regardless of the reason that the IMBHs do not lie on the fundamental plane, the fact that they do not means that simple and straightforward radio and X-ray observations to estimate mass are not warranted.  One needs, at a minimum, to know the Eddington fraction of the source to determine if the fundamental plane is appropriate.} One potential, though flawed, avenue for inferring Eddington fraction is by using $\alpha_{\mathrm{OX}}$.

Despite the implied difficulties in using the fundamental plane as a mass estimator, the fundamental plane still provides interesting and strong empirical data for understanding the energetics of low-Eddington AGN.  Future work to expand the fundamental plane to include the largest supermassive black holes will improve the baseline in mass to better constrain the fundamental plane parameters.

\section*{Acknowledgements}
The National Radio Astronomy Observatory is a facility of the National Science Foundation operated under cooperative agreement by Associated Universities, Inc.  We thank NRAO for providing computing resources for data processing and analysis. 
Basic research in radio astronomy at the U.S. Naval Research Laboratory is supported by 6.1 Base Funding. 
%NRAO for computing resources.

%%%%%%%%%%%%%%%%%%%%%%%%%%%%%%%%%%%%%%%%%%%%%%%%%%
\section*{Data Availability}

The data underlying this article are available in NRAO archive at \url{http://nrao.edu/archive}, and can be accessed with a search for project ID  15A-240.  \kgbf{Images can be accessed in fits format from \url{https://doi.org/10.7302/3100-6e62} or by request to the corresponding author.}



\bibliographystyle{mnras}
%\bibliography{gultekin} % if your bibtex file is called example.bib\
\bibliography{minifp}

\begin{thebibliography}{}
\makeatletter
\relax
\def\mn@urlcharsother{\let\do\@makeother \do\$\do\&\do\#\do\^\do\_\do\%\do\~}
\def\mn@doi{\begingroup\mn@urlcharsother \@ifnextchar [ {\mn@doi@}
  {\mn@doi@[]}}
\def\mn@doi@[#1]#2{\def\@tempa{#1}\ifx\@tempa\@empty \href
  {http://dx.doi.org/#2} {doi:#2}\else \href {http://dx.doi.org/#2} {#1}\fi
  \endgroup}
\def\mn@eprint#1#2{\mn@eprint@#1:#2::\@nil}
\def\mn@eprint@arXiv#1{\href {http://arxiv.org/abs/#1} {{\tt arXiv:#1}}}
\def\mn@eprint@dblp#1{\href {http://dblp.uni-trier.de/rec/bibtex/#1.xml}
  {dblp:#1}}
\def\mn@eprint@#1:#2:#3:#4\@nil{\def\@tempa {#1}\def\@tempb {#2}\def\@tempc
  {#3}\ifx \@tempc \@empty \let \@tempc \@tempb \let \@tempb \@tempa \fi \ifx
  \@tempb \@empty \def\@tempb {arXiv}\fi \@ifundefined
  {mn@eprint@\@tempb}{\@tempb:\@tempc}{\expandafter \expandafter \csname
  mn@eprint@\@tempb\endcsname \expandafter{\@tempc}}}

\bibitem[\protect\citeauthoryear{{Abbott} et~al.,}{{Abbott}
  et~al.}{2020a}]{2020PhRvL.125j1102A}
{Abbott} R.,  et~al., 2020a, \mn@doi [\prl] {10.1103/PhysRevLett.125.101102},
  \href {https://ui.adsabs.harvard.edu/abs/2020PhRvL.125j1102A} {125, 101102}

\bibitem[\protect\citeauthoryear{{Abbott} et~al.,}{{Abbott}
  et~al.}{2020b}]{2020ApJ...900L..13A}
{Abbott} R.,  et~al., 2020b, \mn@doi [\apjl] {10.3847/2041-8213/aba493}, \href
  {https://ui.adsabs.harvard.edu/abs/2020ApJ...900L..13A} {900, L13}

\bibitem[\protect\citeauthoryear{{Amaro-Seoane} et~al.,}{{Amaro-Seoane}
  et~al.}{2017}]{2017arXiv170200786A}
{Amaro-Seoane} P.,  et~al., 2017, arXiv e-prints, \href
  {https://ui.adsabs.harvard.edu/abs/2017arXiv170200786A} {p. arXiv:1702.00786}

\bibitem[\protect\citeauthoryear{{Baldassare}, {Reines}, {Gallo}  \&
  {Greene}}{{Baldassare} et~al.}{2015}]{2015ApJ...809L..14B}
{Baldassare} V.~F.,  {Reines} A.~E.,  {Gallo} E.,   {Greene} J.~E.,  2015,
  \mn@doi [\apjl] {10.1088/2041-8205/809/1/L14}, \href
  {http://adsabs.harvard.edu/abs/2015ApJ...809L..14B} {809, L14}

\bibitem[\protect\citeauthoryear{{Baldassare} et~al.,}{{Baldassare}
  et~al.}{2016}]{2016ApJ...829...57B}
{Baldassare} V.~F.,  et~al., 2016, \mn@doi [\apj] {10.3847/0004-637X/829/1/57},
  \href {https://ui.adsabs.harvard.edu/abs/2016ApJ...829...57B} {829, 57}

\bibitem[\protect\citeauthoryear{{Baldassare}, {Reines}, {Gallo}  \&
  {Greene}}{{Baldassare} et~al.}{2017}]{2017ApJ...836...20B}
{Baldassare} V.~F.,  {Reines} A.~E.,  {Gallo} E.,   {Greene} J.~E.,  2017,
  \mn@doi [\apj] {10.3847/1538-4357/836/1/20}, \href
  {https://ui.adsabs.harvard.edu/abs/2017ApJ...836...20B} {836, 20}

\bibitem[\protect\citeauthoryear{{Baldwin}, {Phillips}  \&
  {Terlevich}}{{Baldwin} et~al.}{1981}]{1981PASP...93....5B}
{Baldwin} J.~A.,  {Phillips} M.~M.,   {Terlevich} R.,  1981, \mn@doi [\pasp]
  {10.1086/130766}, \href
  {https://ui.adsabs.harvard.edu/abs/1981PASP...93....5B} {93, 5}

\bibitem[\protect\citeauthoryear{{Bentz}, {Peterson}, {Pogge}, {Vestergaard}
  \& {Onken}}{{Bentz} et~al.}{2006}]{2006ApJ...644..133B}
{Bentz} M.~C.,  {Peterson} B.~M.,  {Pogge} R.~W.,  {Vestergaard} M.,   {Onken}
  C.~A.,  2006, \mn@doi [\apj] {10.1086/503537}, \href
  {https://ui.adsabs.harvard.edu/abs/2006ApJ...644..133B} {644, 133}

\bibitem[\protect\citeauthoryear{{Bentz}, {Peterson}, {Netzer}, {Pogge}  \&
  {Vestergaard}}{{Bentz} et~al.}{2009}]{2009ApJ...697..160B}
{Bentz} M.~C.,  {Peterson} B.~M.,  {Netzer} H.,  {Pogge} R.~W.,   {Vestergaard}
  M.,  2009, \mn@doi [\apj] {10.1088/0004-637X/697/1/160}, \href
  {https://ui.adsabs.harvard.edu/abs/2009ApJ...697..160B} {697, 160}

\bibitem[\protect\citeauthoryear{{Bentz} et~al.,}{{Bentz}
  et~al.}{2013}]{2013ApJ...767..149B}
{Bentz} M.~C.,  et~al., 2013, \mn@doi [\apj] {10.1088/0004-637X/767/2/149},
  \href {https://ui.adsabs.harvard.edu/abs/2013ApJ...767..149B} {767, 149}

\bibitem[\protect\citeauthoryear{{Bentz}, {Williams}  \& {Treu}}{{Bentz}
  et~al.}{2022}]{2022arXiv220603513B}
{Bentz} M.~C.,  {Williams} P.~R.,   {Treu} T.,  2022, arXiv e-prints, \href
  {https://ui.adsabs.harvard.edu/abs/2022arXiv220603513B} {p. arXiv:2206.03513}

\bibitem[\protect\citeauthoryear{{Berton}, {Foschini}, {Ciroi}, {Cracco}, {La
  Mura}, {Di Mille}  \& {Rafanelli}}{{Berton}
  et~al.}{2016}]{2016A&A...591A..88B}
{Berton} M.,  {Foschini} L.,  {Ciroi} S.,  {Cracco} V.,  {La Mura} G.,  {Di
  Mille} F.,   {Rafanelli} P.,  2016, \mn@doi [\aap]
  {10.1051/0004-6361/201527056}, \href
  {https://ui.adsabs.harvard.edu/abs/2016A&A...591A..88B} {591, A88}

\bibitem[\protect\citeauthoryear{{Brooks} \& {Christensen}}{{Brooks} \&
  {Christensen}}{2016}]{2016ASSL..418..317B}
{Brooks} A.,  {Christensen} C.,  2016, in {Laurikainen} E.,  {Peletier} R.,
  {Gadotti} D.,  eds,  Astrophysics and Space Science Library Vol. 418,
  Galactic Bulges. p.~317 (\mn@eprint {arXiv} {1511.04095}),
  \mn@doi{10.1007/978-3-319-19378-6\_12}

\bibitem[\protect\citeauthoryear{{Carilli} et~al.,}{{Carilli}
  et~al.}{2015}]{2015arXiv151006438C}
{Carilli} C.~L.,  et~al., 2015, arXiv e-prints, \href
  {https://ui.adsabs.harvard.edu/abs/2015arXiv151006438C} {p. arXiv:1510.06438}

\bibitem[\protect\citeauthoryear{{Casey} et~al.,}{{Casey}
  et~al.}{2015}]{2015arXiv151006411C}
{Casey} C.~M.,  et~al., 2015, arXiv e-prints, \href
  {https://ui.adsabs.harvard.edu/abs/2015arXiv151006411C} {p. arXiv:1510.06411}

\bibitem[\protect\citeauthoryear{{Chen} et~al.,}{{Chen}
  et~al.}{2020}]{2020ApJ...897..102C}
{Chen} Z.,  et~al., 2020, \mn@doi [\apj] {10.3847/1538-4357/ab9633}, \href
  {https://ui.adsabs.harvard.edu/abs/2020ApJ...897..102C} {897, 102}

\bibitem[\protect\citeauthoryear{{Davis} et~al.,}{{Davis}
  et~al.}{2020}]{2020MNRAS.496.4061D}
{Davis} T.~A.,  et~al., 2020, \mn@doi [\mnras] {10.1093/mnras/staa1567}, \href
  {https://ui.adsabs.harvard.edu/abs/2020MNRAS.496.4061D} {496, 4061}

\bibitem[\protect\citeauthoryear{{Dickey}, {Geha}, {Wetzel}  \&
  {El-Badry}}{{Dickey} et~al.}{2019}]{2019ApJ...884..180D}
{Dickey} C.~M.,  {Geha} M.,  {Wetzel} A.,   {El-Badry} K.,  2019, \mn@doi
  [\apj] {10.3847/1538-4357/ab3220}, \href
  {https://ui.adsabs.harvard.edu/abs/2019ApJ...884..180D} {884, 180}

\bibitem[\protect\citeauthoryear{{Dong}, {Greene}  \& {Ho}}{{Dong}
  et~al.}{2012}]{2012ApJ...761...73D}
{Dong} R.,  {Greene} J.~E.,   {Ho} L.~C.,  2012, \mn@doi [\apj]
  {10.1088/0004-637X/761/1/73}, \href
  {https://ui.adsabs.harvard.edu/abs/2012ApJ...761...73D} {761, 73}

\bibitem[\protect\citeauthoryear{{Dong}, {Wu}  \& {Cao}}{{Dong}
  et~al.}{2014}]{2014ApJ...787L..20D}
{Dong} A.-J.,  {Wu} Q.,   {Cao} X.-F.,  2014, \mn@doi [\apjl]
  {10.1088/2041-8205/787/2/L20}, \href
  {https://ui.adsabs.harvard.edu/abs/2014ApJ...787L..20D} {787, L20}

\bibitem[\protect\citeauthoryear{{Eracleous}, {Hwang}  \& {Flohic}}{{Eracleous}
  et~al.}{2010}]{2010ApJS..187..135E}
{Eracleous} M.,  {Hwang} J.~A.,   {Flohic} H. M.~L.~G.,  2010, \mn@doi [\apjs]
  {10.1088/0067-0049/187/1/135}, \href
  {https://ui.adsabs.harvard.edu/abs/2010ApJS..187..135E} {187, 135}

\bibitem[\protect\citeauthoryear{{Falcke}, {K{\"o}rding}  \&
  {Markoff}}{{Falcke} et~al.}{2004}]{2004A&A...414..895F}
{Falcke} H.,  {K{\"o}rding} E.,   {Markoff} S.,  2004, \mn@doi [\aap]
  {10.1051/0004-6361:20031683}, \href
  {https://ui.adsabs.harvard.edu/abs/2004A&A...414..895F} {414, 895}

\bibitem[\protect\citeauthoryear{{Fan} \& {Bai}}{{Fan} \&
  {Bai}}{2016}]{2016ApJ...818..185F}
{Fan} X.-L.,  {Bai} J.-M.,  2016, \mn@doi [\apj] {10.3847/0004-637X/818/2/185},
  \href {https://ui.adsabs.harvard.edu/abs/2016ApJ...818..185F} {818, 185}

\bibitem[\protect\citeauthoryear{{Fender}, {Belloni}  \& {Gallo}}{{Fender}
  et~al.}{2004}]{2004MNRAS.355.1105F}
{Fender} R.~P.,  {Belloni} T.~M.,   {Gallo} E.,  2004, \mn@doi [\mnras]
  {10.1111/j.1365-2966.2004.08384.x}, \href
  {https://ui.adsabs.harvard.edu/abs/2004MNRAS.355.1105F} {355, 1105}

\bibitem[\protect\citeauthoryear{{Fischer} et~al.,}{{Fischer}
  et~al.}{2021}]{2021ApJ...906...88F}
{Fischer} T.~C.,  et~al., 2021, \mn@doi [\apj] {10.3847/1538-4357/abca3c},
  \href {https://ui.adsabs.harvard.edu/abs/2021ApJ...906...88F} {906, 88}

\bibitem[\protect\citeauthoryear{{Fisher} \& {Drory}}{{Fisher} \&
  {Drory}}{2010}]{2010ApJ...716..942F}
{Fisher} D.~B.,  {Drory} N.,  2010, \mn@doi [\apj]
  {10.1088/0004-637X/716/2/942}, \href
  {https://ui.adsabs.harvard.edu/abs/2010ApJ...716..942F} {716, 942}

\bibitem[\protect\citeauthoryear{{Fisher} \& {Drory}}{{Fisher} \&
  {Drory}}{2011}]{2011ApJ...733L..47F}
{Fisher} D.~B.,  {Drory} N.,  2011, \mn@doi [\apjl]
  {10.1088/2041-8205/733/2/L47}, \href
  {https://ui.adsabs.harvard.edu/abs/2011ApJ...733L..47F} {733, L47}

\bibitem[\protect\citeauthoryear{{Fryer}, {Woosley}  \& {Heger}}{{Fryer}
  et~al.}{2001}]{2001ApJ...550..372F}
{Fryer} C.~L.,  {Woosley} S.~E.,   {Heger} A.,  2001, \mn@doi [\apj]
  {10.1086/319719}, \href
  {https://ui.adsabs.harvard.edu/abs/2001ApJ...550..372F} {550, 372}

\bibitem[\protect\citeauthoryear{{Gallo}, {Fender}  \& {Pooley}}{{Gallo}
  et~al.}{2003}]{2003MNRAS.344...60G}
{Gallo} E.,  {Fender} R.~P.,   {Pooley} G.~G.,  2003, \mn@doi [\mnras]
  {10.1046/j.1365-8711.2003.06791.x}, \href
  {https://ui.adsabs.harvard.edu/abs/2003MNRAS.344...60G} {344, 60}

\bibitem[\protect\citeauthoryear{{Gallo}, {Miller}  \& {Fender}}{{Gallo}
  et~al.}{2012}]{2012MNRAS.423..590G}
{Gallo} E.,  {Miller} B.~P.,   {Fender} R.,  2012, \mn@doi [\mnras]
  {10.1111/j.1365-2966.2012.20899.x}, \href
  {https://ui.adsabs.harvard.edu/abs/2012MNRAS.423..590G} {423, 590}

\bibitem[\protect\citeauthoryear{{Greene} \& {Ho}}{{Greene} \&
  {Ho}}{2007}]{2007ApJ...656...84G}
{Greene} J.~E.,  {Ho} L.~C.,  2007, \mn@doi [\apj] {10.1086/509064}, \href
  {https://ui.adsabs.harvard.edu/abs/2007ApJ...656...84G} {656, 84}

\bibitem[\protect\citeauthoryear{{Greene}, {Ho}  \& {Ulvestad}}{{Greene}
  et~al.}{2006}]{2006ApJ...636...56G}
{Greene} J.~E.,  {Ho} L.~C.,   {Ulvestad} J.~S.,  2006, \mn@doi [\apj]
  {10.1086/497905}, \href
  {https://ui.adsabs.harvard.edu/abs/2006ApJ...636...56G} {636, 56}

\bibitem[\protect\citeauthoryear{{Greene}, {Ho}  \& {Barth}}{{Greene}
  et~al.}{2008}]{2008ApJ...688..159G}
{Greene} J.~E.,  {Ho} L.~C.,   {Barth} A.~J.,  2008, \mn@doi [\apj]
  {10.1086/592078}, \href
  {https://ui.adsabs.harvard.edu/abs/2008ApJ...688..159G} {688, 159}

\bibitem[\protect\citeauthoryear{{Greene}, {Strader}  \& {Ho}}{{Greene}
  et~al.}{2020}]{2020ARA&A..58..257G}
{Greene} J.~E.,  {Strader} J.,   {Ho} L.~C.,  2020, \mn@doi [\araa]
  {10.1146/annurev-astro-032620-021835}, \href
  {https://ui.adsabs.harvard.edu/abs/2020ARA&A..58..257G} {58, 257}

\bibitem[\protect\citeauthoryear{{G{\"u}ltekin}, {Miller}  \&
  {Hamilton}}{{G{\"u}ltekin} et~al.}{2004}]{2004ApJ...616..221G}
{G{\"u}ltekin} K.,  {Miller} M.~C.,   {Hamilton} D.~P.,  2004, \mn@doi [\apj]
  {10.1086/424809}, \href
  {https://ui.adsabs.harvard.edu/abs/2004ApJ...616..221G} {616, 221}

\bibitem[\protect\citeauthoryear{{G{\"u}ltekin}, {Miller}  \&
  {Hamilton}}{{G{\"u}ltekin} et~al.}{2006}]{2006ApJ...640..156G}
{G{\"u}ltekin} K.,  {Miller} M.~C.,   {Hamilton} D.~P.,  2006, \mn@doi [\apj]
  {10.1086/499917}, \href
  {https://ui.adsabs.harvard.edu/abs/2006ApJ...640..156G} {640, 156}

\bibitem[\protect\citeauthoryear{{G{\"u}ltekin} et~al.,}{{G{\"u}ltekin}
  et~al.}{2009a}]{2009ApJ...698..198G}
{G{\"u}ltekin} K.,  et~al., 2009a, \mn@doi [\apj]
  {10.1088/0004-637X/698/1/198}, \href
  {https://ui.adsabs.harvard.edu/abs/2009ApJ...698..198G} {698, 198}

\bibitem[\protect\citeauthoryear{{G{\"u}ltekin}, {Cackett}, {Miller}, {Di
  Matteo}, {Markoff}  \& {Richstone}}{{G{\"u}ltekin}
  et~al.}{2009b}]{2009ApJ...706..404G}
{G{\"u}ltekin} K.,  {Cackett} E.~M.,  {Miller} J.~M.,  {Di Matteo} T.,
  {Markoff} S.,   {Richstone} D.~O.,  2009b, \mn@doi [\apj]
  {10.1088/0004-637X/706/1/404}, \href
  {https://ui.adsabs.harvard.edu/abs/2009ApJ...706..404G} {706, 404}

\bibitem[\protect\citeauthoryear{{G{\"u}ltekin}, {Cackett}, {Miller}, {Di
  Matteo}, {Markoff}  \& {Richstone}}{{G{\"u}ltekin}
  et~al.}{2012}]{2012ApJ...749..129G}
{G{\"u}ltekin} K.,  {Cackett} E.~M.,  {Miller} J.~M.,  {Di Matteo} T.,
  {Markoff} S.,   {Richstone} D.~O.,  2012, \mn@doi [\apj]
  {10.1088/0004-637X/749/2/129}, \href
  {https://ui.adsabs.harvard.edu/abs/2012ApJ...749..129G} {749, 129}

\bibitem[\protect\citeauthoryear{{G{\"u}ltekin}, {Cackett}, {King}, {Miller}
  \& {Pinkney}}{{G{\"u}ltekin} et~al.}{2014}]{2014ApJ...788L..22G}
{G{\"u}ltekin} K.,  {Cackett} E.~M.,  {King} A.~L.,  {Miller} J.~M.,
  {Pinkney} J.,  2014, \mn@doi [\apjl] {10.1088/2041-8205/788/2/L22}, \href
  {https://ui.adsabs.harvard.edu/abs/2014ApJ...788L..22G} {788, L22}

\bibitem[\protect\citeauthoryear{{G{\"u}ltekin}, {King}, {Cackett}, {Nyland},
  {Miller}, {Di Matteo}, {Markoff}  \& {Rupen}}{{G{\"u}ltekin}
  et~al.}{2019}]{2019ApJ...871...80G}
{G{\"u}ltekin} K.,  {King} A.~L.,  {Cackett} E.~M.,  {Nyland} K.,  {Miller}
  J.~M.,  {Di Matteo} T.,  {Markoff} S.,   {Rupen} M.~P.,  2019, \mn@doi [\apj]
  {10.3847/1538-4357/aaf6b9}, \href
  {https://ui.adsabs.harvard.edu/abs/2019ApJ...871...80G} {871, 80}

\bibitem[\protect\citeauthoryear{{Hallinan} et~al.,}{{Hallinan}
  et~al.}{2019}]{2019BAAS...51g.255H}
{Hallinan} G.,  et~al., 2019, in Bulletin of the American Astronomical Society.
  p.~255 (\mn@eprint {arXiv} {1907.07648})

\bibitem[\protect\citeauthoryear{{Heinz} \& {Sunyaev}}{{Heinz} \&
  {Sunyaev}}{2003}]{2003MNRAS.343L..59H}
{Heinz} S.,  {Sunyaev} R.~A.,  2003, \mn@doi [\mnras]
  {10.1046/j.1365-8711.2003.06918.x}, \href
  {https://ui.adsabs.harvard.edu/abs/2003MNRAS.343L..59H} {343, L59}

\bibitem[\protect\citeauthoryear{{Ho}}{{Ho}}{1999}]{1999ApJ...516..672H}
{Ho} L.~C.,  1999, \mn@doi [\apj] {10.1086/307137}, \href
  {https://ui.adsabs.harvard.edu/abs/1999ApJ...516..672H} {516, 672}

\bibitem[\protect\citeauthoryear{{Ho} \& {Kim}}{{Ho} \&
  {Kim}}{2014}]{2014ApJ...789...17H}
{Ho} L.~C.,  {Kim} M.,  2014, \mn@doi [\apj] {10.1088/0004-637X/789/1/17},
  \href {https://ui.adsabs.harvard.edu/abs/2014ApJ...789...17H} {789, 17}

\bibitem[\protect\citeauthoryear{{Kewley}, {Groves}, {Kauffmann}  \&
  {Heckman}}{{Kewley} et~al.}{2006}]{2006MNRAS.372..961K}
{Kewley} L.~J.,  {Groves} B.,  {Kauffmann} G.,   {Heckman} T.,  2006, \mn@doi
  [\mnras] {10.1111/j.1365-2966.2006.10859.x}, \href
  {https://ui.adsabs.harvard.edu/abs/2006MNRAS.372..961K} {372, 961}

\bibitem[\protect\citeauthoryear{{Kormendy} \& {Ho}}{{Kormendy} \&
  {Ho}}{2013}]{2013ARA&A..51..511K}
{Kormendy} J.,  {Ho} L.~C.,  2013, \mn@doi [\araa]
  {10.1146/annurev-astro-082708-101811}, \href
  {https://ui.adsabs.harvard.edu/abs/2013ARA&A..51..511K} {51, 511}

\bibitem[\protect\citeauthoryear{{Kormendy} \& {Kennicutt}}{{Kormendy} \&
  {Kennicutt}}{2004}]{2004ARA&A..42..603K}
{Kormendy} J.,  {Kennicutt} Robert~C. J.,  2004, \mn@doi [\araa]
  {10.1146/annurev.astro.42.053102.134024}, \href
  {https://ui.adsabs.harvard.edu/abs/2004ARA&A..42..603K} {42, 603}

\bibitem[\protect\citeauthoryear{{Leighly}, {Halpern}, {Jenkins}, {Grupe},
  {Choi}  \& {Prescott}}{{Leighly} et~al.}{2007}]{2007ApJ...663..103L}
{Leighly} K.~M.,  {Halpern} J.~P.,  {Jenkins} E.~B.,  {Grupe} D.,  {Choi} J.,
  {Prescott} K.~B.,  2007, \mn@doi [\apj] {10.1086/518017}, \href
  {https://ui.adsabs.harvard.edu/abs/2007ApJ...663..103L} {663, 103}

\bibitem[\protect\citeauthoryear{{Li}, {Wu}  \& {Wang}}{{Li}
  et~al.}{2008}]{2008ApJ...688..826L}
{Li} Z.-Y.,  {Wu} X.-B.,   {Wang} R.,  2008, \mn@doi [\apj] {10.1086/592314},
  \href {https://ui.adsabs.harvard.edu/abs/2008ApJ...688..826L} {688, 826}

\bibitem[\protect\citeauthoryear{{Luo} et~al.,}{{Luo}
  et~al.}{2015}]{2015ApJ...805..122L}
{Luo} B.,  et~al., 2015, \mn@doi [\apj] {10.1088/0004-637X/805/2/122}, \href
  {https://ui.adsabs.harvard.edu/abs/2015ApJ...805..122L} {805, 122}

\bibitem[\protect\citeauthoryear{{Luo} et~al.,}{{Luo}
  et~al.}{2020}]{2020MNRAS.493.1686L}
{Luo} Y.,  et~al., 2020, \mn@doi [\mnras] {10.1093/mnras/staa328}, \href
  {https://ui.adsabs.harvard.edu/abs/2020MNRAS.493.1686L} {493, 1686}

\bibitem[\protect\citeauthoryear{{Lusso} et~al.,}{{Lusso}
  et~al.}{2010}]{2010A&A...512A..34L}
{Lusso} E.,  et~al., 2010, \mn@doi [\aap] {10.1051/0004-6361/200913298}, \href
  {https://ui.adsabs.harvard.edu/abs/2010A&A...512A..34L} {512, A34}

\bibitem[\protect\citeauthoryear{{MacLeod}, {Guillochon}, {Ramirez-Ruiz},
  {Kasen}  \& {Rosswog}}{{MacLeod} et~al.}{2016}]{2016ApJ...819....3M}
{MacLeod} M.,  {Guillochon} J.,  {Ramirez-Ruiz} E.,  {Kasen} D.,   {Rosswog}
  S.,  2016, \mn@doi [\apj] {10.3847/0004-637X/819/1/3}, \href
  {https://ui.adsabs.harvard.edu/abs/2016ApJ...819....3M} {819, 3}

\bibitem[\protect\citeauthoryear{{Maccarone}, {Gallo}  \& {Fender}}{{Maccarone}
  et~al.}{2003}]{2003MNRAS.345L..19M}
{Maccarone} T.~J.,  {Gallo} E.,   {Fender} R.,  2003, \mn@doi [\mnras]
  {10.1046/j.1365-8711.2003.07161.x}, \href
  {https://ui.adsabs.harvard.edu/abs/2003MNRAS.345L..19M} {345, L19}

\bibitem[\protect\citeauthoryear{{Madau} \& {Haardt}}{{Madau} \&
  {Haardt}}{2015}]{2015ApJ...813L...8M}
{Madau} P.,  {Haardt} F.,  2015, \mn@doi [\apjl] {10.1088/2041-8205/813/1/L8},
  \href {https://ui.adsabs.harvard.edu/abs/2015ApJ...813L...8M} {813, L8}

\bibitem[\protect\citeauthoryear{{Madau} \& {Rees}}{{Madau} \&
  {Rees}}{2001}]{2001ApJ...551L..27M}
{Madau} P.,  {Rees} M.~J.,  2001, \mn@doi [\apjl] {10.1086/319848}, \href
  {https://ui.adsabs.harvard.edu/abs/2001ApJ...551L..27M} {551, L27}

\bibitem[\protect\citeauthoryear{{Maitra}, {Miller}, {Markoff}  \&
  {King}}{{Maitra} et~al.}{2011}]{2011ApJ...735..107M}
{Maitra} D.,  {Miller} J.~M.,  {Markoff} S.,   {King} A.,  2011, \mn@doi [\apj]
  {10.1088/0004-637X/735/2/107}, \href
  {https://ui.adsabs.harvard.edu/abs/2011ApJ...735..107M} {735, 107}

\bibitem[\protect\citeauthoryear{{Marchese}, {Della Ceca}, {Caccianiga},
  {Severgnini}, {Corral}  \& {Fanali}}{{Marchese}
  et~al.}{2012}]{2012A&A...539A..48M}
{Marchese} E.,  {Della Ceca} R.,  {Caccianiga} A.,  {Severgnini} P.,  {Corral}
  A.,   {Fanali} R.,  2012, \mn@doi [\aap] {10.1051/0004-6361/201117562}, \href
  {https://ui.adsabs.harvard.edu/abs/2012A&A...539A..48M} {539, A48}

\bibitem[\protect\citeauthoryear{{Marconi}, {Risaliti}, {Gilli}, {Hunt},
  {Maiolino}  \& {Salvati}}{{Marconi} et~al.}{2004}]{2004MNRAS.351..169M}
{Marconi} A.,  {Risaliti} G.,  {Gilli} R.,  {Hunt} L.~K.,  {Maiolino} R.,
  {Salvati} M.,  2004, \mn@doi [\mnras] {10.1111/j.1365-2966.2004.07765.x},
  \href {https://ui.adsabs.harvard.edu/abs/2004MNRAS.351..169M} {351, 169}

\bibitem[\protect\citeauthoryear{{Markoff}, {Falcke}, {Yuan}  \&
  {Biermann}}{{Markoff} et~al.}{2001}]{2001A&A...379L..13M}
{Markoff} S.,  {Falcke} H.,  {Yuan} F.,   {Biermann} P.~L.,  2001, \mn@doi
  [\aap] {10.1051/0004-6361:20011346}, \href
  {https://ui.adsabs.harvard.edu/abs/2001A&A...379L..13M} {379, L13}

\bibitem[\protect\citeauthoryear{{McHardy}, {Koerding}, {Knigge}, {Uttley}  \&
  {Fender}}{{McHardy} et~al.}{2006}]{2006Natur.444..730M}
{McHardy} I.~M.,  {Koerding} E.,  {Knigge} C.,  {Uttley} P.,   {Fender} R.~P.,
  2006, \mn@doi [\nat] {10.1038/nature05389}, \href
  {https://ui.adsabs.harvard.edu/abs/2006Natur.444..730M} {444, 730}

\bibitem[\protect\citeauthoryear{{McKinnon}, {Carilli}  \&
  {Beasley}}{{McKinnon} et~al.}{2016}]{2016SPIE.9906E..27M}
{McKinnon} M.,  {Carilli} C.,   {Beasley} T.,  2016, in {Hall} H.~J.,
  {Gilmozzi} R.,   {Marshall} H.~K.,  eds,  Society of Photo-Optical
  Instrumentation Engineers (SPIE) Conference Series Vol. 9906, Ground-based
  and Airborne Telescopes VI. p. 990627, \mn@doi{10.1117/12.2234941}

\bibitem[\protect\citeauthoryear{{McMullin}, {Waters}, {Schiebel}, {Young}  \&
  {Golap}}{{McMullin} et~al.}{2007}]{2007ASPC..376..127M}
{McMullin} J.~P.,  {Waters} B.,  {Schiebel} D.,  {Young} W.,   {Golap} K.,
  2007, in {Shaw} R.~A.,  {Hill} F.,   {Bell} D.~J.,  eds,  Astronomical
  Society of the Pacific Conference Series Vol. 376, Astronomical Data Analysis
  Software and Systems XVI. p.~127

\bibitem[\protect\citeauthoryear{{Merloni} \& {Heinz}}{{Merloni} \&
  {Heinz}}{2007}]{2007MNRAS.381..589M}
{Merloni} A.,  {Heinz} S.,  2007, \mn@doi [\mnras]
  {10.1111/j.1365-2966.2007.12253.x}, \href
  {https://ui.adsabs.harvard.edu/abs/2007MNRAS.381..589M} {381, 589}

\bibitem[\protect\citeauthoryear{{Merloni}, {Heinz}  \& {di Matteo}}{{Merloni}
  et~al.}{2003}]{2003MNRAS.345.1057M}
{Merloni} A.,  {Heinz} S.,   {di Matteo} T.,  2003, \mn@doi [\mnras]
  {10.1046/j.1365-2966.2003.07017.x}, \href
  {https://ui.adsabs.harvard.edu/abs/2003MNRAS.345.1057M} {345, 1057}

\bibitem[\protect\citeauthoryear{{Merloni}, {K{\"o}rding}, {Heinz}, {Markoff},
  {Di Matteo}  \& {Falcke}}{{Merloni} et~al.}{2006}]{2006NewA...11..567M}
{Merloni} A.,  {K{\"o}rding} E.,  {Heinz} S.,  {Markoff} S.,  {Di Matteo} T.,
  {Falcke} H.,  2006, \mn@doi [\na] {10.1016/j.newast.2006.03.002}, \href
  {https://ui.adsabs.harvard.edu/abs/2006NewA...11..567M} {11, 567}

\bibitem[\protect\citeauthoryear{{Mezcua}, {Suh}  \& {Civano}}{{Mezcua}
  et~al.}{2019}]{2019MNRAS.488..685M}
{Mezcua} M.,  {Suh} H.,   {Civano} F.,  2019, \mn@doi [\mnras]
  {10.1093/mnras/stz1760}, \href
  {https://ui.adsabs.harvard.edu/abs/2019MNRAS.488..685M} {488, 685}

\bibitem[\protect\citeauthoryear{{Miller} \& {Hamilton}}{{Miller} \&
  {Hamilton}}{2002}]{2002MNRAS.330..232C}
{Miller} M.~C.,  {Hamilton} D.~P.,  2002, \mn@doi [\mnras]
  {10.1046/j.1365-8711.2002.05112.x}, \href
  {https://ui.adsabs.harvard.edu/abs/2002MNRAS.330..232C} {330, 232}

\bibitem[\protect\citeauthoryear{{Molina}, {Reines}, {Greene}, {Darling}  \&
  {Condon}}{{Molina} et~al.}{2021}]{2021ApJ...910....5M}
{Molina} M.,  {Reines} A.~E.,  {Greene} J.~E.,  {Darling} J.,   {Condon} J.~J.,
   2021, \mn@doi [\apj] {10.3847/1538-4357/abe120}, \href
  {https://ui.adsabs.harvard.edu/abs/2021ApJ...910....5M} {910, 5}

\bibitem[\protect\citeauthoryear{{Nemmen}, {Storchi-Bergmann}  \&
  {Eracleous}}{{Nemmen} et~al.}{2014}]{2014MNRAS.438.2804N}
{Nemmen} R.~S.,  {Storchi-Bergmann} T.,   {Eracleous} M.,  2014, \mn@doi
  [\mnras] {10.1093/mnras/stt2388}, \href
  {https://ui.adsabs.harvard.edu/abs/2014MNRAS.438.2804N} {438, 2804}

\bibitem[\protect\citeauthoryear{{Nguyen} et~al.,}{{Nguyen}
  et~al.}{2017}]{2017ApJ...836..237N}
{Nguyen} D.~D.,  et~al., 2017, \mn@doi [\apj] {10.3847/1538-4357/aa5cb4}, \href
  {https://ui.adsabs.harvard.edu/abs/2017ApJ...836..237N} {836, 237}

\bibitem[\protect\citeauthoryear{{Nguyen} et~al.,}{{Nguyen}
  et~al.}{2018}]{2018ApJ...858..118N}
{Nguyen} D.~D.,  et~al., 2018, \mn@doi [\apj] {10.3847/1538-4357/aabe28}, \href
  {https://ui.adsabs.harvard.edu/abs/2018ApJ...858..118N} {858, 118}

\bibitem[\protect\citeauthoryear{{Nyland} et~al.,}{{Nyland}
  et~al.}{2017}]{nyland+17}
{Nyland} K.,  et~al., 2017, \mn@doi [\apj] {10.3847/1538-4357/aa7ecf}, \href
  {https://ui.adsabs.harvard.edu/abs/2017ApJ...845...50N} {845, 50}

\bibitem[\protect\citeauthoryear{{Nyland} et~al.,}{{Nyland}
  et~al.}{2018}]{nyland+18}
{Nyland} K.,  et~al., 2018, \mn@doi [\apj] {10.3847/1538-4357/aab3d1}, \href
  {https://ui.adsabs.harvard.edu/abs/2018ApJ...859...23N} {859, 23}

\bibitem[\protect\citeauthoryear{{Nyland} et~al.,}{{Nyland}
  et~al.}{2020}]{2020ApJ...905...74N}
{Nyland} K.,  et~al., 2020, \mn@doi [\apj] {10.3847/1538-4357/abc341}, \href
  {https://ui.adsabs.harvard.edu/abs/2020ApJ...905...74N} {905, 74}

\bibitem[\protect\citeauthoryear{{Pancoast}, {Brewer}  \& {Treu}}{{Pancoast}
  et~al.}{2011}]{2011ApJ...730..139P}
{Pancoast} A.,  {Brewer} B.~J.,   {Treu} T.,  2011, \mn@doi [\apj]
  {10.1088/0004-637X/730/2/139}, \href
  {https://ui.adsabs.harvard.edu/abs/2011ApJ...730..139P} {730, 139}

\bibitem[\protect\citeauthoryear{{Pancoast}, {Brewer}  \& {Treu}}{{Pancoast}
  et~al.}{2014}]{2014MNRAS.445.3055P}
{Pancoast} A.,  {Brewer} B.~J.,   {Treu} T.,  2014, \mn@doi [\mnras]
  {10.1093/mnras/stu1809}, \href
  {https://ui.adsabs.harvard.edu/abs/2014MNRAS.445.3055P} {445, 3055}

\bibitem[\protect\citeauthoryear{{Parker}, {Alston}, {Igo}  \&
  {Fabian}}{{Parker} et~al.}{2020}]{2020MNRAS.492.1363P}
{Parker} M.~L.,  {Alston} W.~N.,  {Igo} Z.,   {Fabian} A.~C.,  2020, \mn@doi
  [\mnras] {10.1093/mnras/stz3470}, \href
  {https://ui.adsabs.harvard.edu/abs/2020MNRAS.492.1363P} {492, 1363}

\bibitem[\protect\citeauthoryear{{Perley} \& {Butler}}{{Perley} \&
  {Butler}}{2017}]{2017ApJS..230....7P}
{Perley} R.~A.,  {Butler} B.~J.,  2017, \mn@doi [\apjs]
  {10.3847/1538-4365/aa6df9}, \href
  {https://ui.adsabs.harvard.edu/abs/2017ApJS..230....7P} {230, 7}

\bibitem[\protect\citeauthoryear{{Peterson}}{{Peterson}}{2014}]{2014SSRv..183..253P}
{Peterson} B.~M.,  2014, \mn@doi [\ssr] {10.1007/s11214-013-9987-4}, \href
  {https://ui.adsabs.harvard.edu/abs/2014SSRv..183..253P} {183, 253}

\bibitem[\protect\citeauthoryear{{Plotkin}, {Markoff}, {Kelly}, {K{\"o}rding}
  \& {Anderson}}{{Plotkin} et~al.}{2012}]{2012MNRAS.419..267P}
{Plotkin} R.~M.,  {Markoff} S.,  {Kelly} B.~C.,  {K{\"o}rding} E.,   {Anderson}
  S.~F.,  2012, \mn@doi [\mnras] {10.1111/j.1365-2966.2011.19689.x}, \href
  {https://ui.adsabs.harvard.edu/abs/2012MNRAS.419..267P} {419, 267}

\bibitem[\protect\citeauthoryear{{Plotkin}, {Gallo}  \& {Jonker}}{{Plotkin}
  et~al.}{2013}]{2013ApJ...773...59P}
{Plotkin} R.~M.,  {Gallo} E.,   {Jonker} P.~G.,  2013, \mn@doi [\apj]
  {10.1088/0004-637X/773/1/59}, \href
  {https://ui.adsabs.harvard.edu/abs/2013ApJ...773...59P} {773, 59}

\bibitem[\protect\citeauthoryear{{Qian}, {Dong}, {Xie}, {Liu}  \& {Li}}{{Qian}
  et~al.}{2018}]{2018ApJ...860..134Q}
{Qian} L.,  {Dong} X.-B.,  {Xie} F.-G.,  {Liu} W.,   {Li} D.,  2018, \mn@doi
  [\apj] {10.3847/1538-4357/aac32b}, \href
  {https://ui.adsabs.harvard.edu/abs/2018ApJ...860..134Q} {860, 134}

\bibitem[\protect\citeauthoryear{{Reines}}{{Reines}}{2022}]{Reines2022}
{Reines} A.~E.,  2022, \mn@doi [Nature Astronomy] {10.1038/s41550-021-01556-0},
  \href {https://ui.adsabs.harvard.edu/abs/2022NatAs...6...26R} {6, 26}

\bibitem[\protect\citeauthoryear{{Reines}, {Greene}  \& {Geha}}{{Reines}
  et~al.}{2013}]{2013ApJ...775..116R}
{Reines} A.~E.,  {Greene} J.~E.,   {Geha} M.,  2013, \mn@doi [\apj]
  {10.1088/0004-637X/775/2/116}, \href
  {https://ui.adsabs.harvard.edu/abs/2013ApJ...775..116R} {775, 116}

\bibitem[\protect\citeauthoryear{{Reines}, {Condon}, {Darling}  \&
  {Greene}}{{Reines} et~al.}{2020}]{2020ApJ...888...36R}
{Reines} A.~E.,  {Condon} J.~J.,  {Darling} J.,   {Greene} J.~E.,  2020,
  \mn@doi [\apj] {10.3847/1538-4357/ab4999}, \href
  {https://ui.adsabs.harvard.edu/abs/2020ApJ...888...36R} {888, 36}

\bibitem[\protect\citeauthoryear{{Remillard} \& {McClintock}}{{Remillard} \&
  {McClintock}}{2006}]{2006ARA&A..44...49R}
{Remillard} R.~A.,  {McClintock} J.~E.,  2006, \mn@doi [\araa]
  {10.1146/annurev.astro.44.051905.092532}, \href
  {https://ui.adsabs.harvard.edu/abs/2006ARA&A..44...49R} {44, 49}

\bibitem[\protect\citeauthoryear{{Romero}, {Boettcher}, {Markoff}  \&
  {Tavecchio}}{{Romero} et~al.}{2017}]{2017SSRv..207....5R}
{Romero} G.~E.,  {Boettcher} M.,  {Markoff} S.,   {Tavecchio} F.,  2017,
  \mn@doi [\ssr] {10.1007/s11214-016-0328-2}, \href
  {https://ui.adsabs.harvard.edu/abs/2017SSRv..207....5R} {207, 5}

\bibitem[\protect\citeauthoryear{{Runnoe}, {Brotherton}  \& {Shang}}{{Runnoe}
  et~al.}{2012}]{2012MNRAS.422..478R}
{Runnoe} J.~C.,  {Brotherton} M.~S.,   {Shang} Z.,  2012, \mn@doi [\mnras]
  {10.1111/j.1365-2966.2012.20620.x}, \href
  {https://ui.adsabs.harvard.edu/abs/2012MNRAS.422..478R} {422, 478}

\bibitem[\protect\citeauthoryear{{Schutte} \& {Reines}}{{Schutte} \&
  {Reines}}{2022}]{2022Natur.601..329S}
{Schutte} Z.,  {Reines} A.~E.,  2022, \mn@doi [\nat]
  {10.1038/s41586-021-04215-6}, \href
  {https://ui.adsabs.harvard.edu/abs/2022Natur.601..329S} {601, 329}

\bibitem[\protect\citeauthoryear{{Seth} et~al.,}{{Seth}
  et~al.}{2010}]{2010ApJ...714..713S}
{Seth} A.~C.,  et~al., 2010, \mn@doi [\apj] {10.1088/0004-637X/714/1/713},
  \href {https://ui.adsabs.harvard.edu/abs/2010ApJ...714..713S} {714, 713}

\bibitem[\protect\citeauthoryear{{Shahbaz}, {Russell}, {Zurita}, {Casares},
  {Corral-Santana}, {Dhillon}  \& {Marsh}}{{Shahbaz}
  et~al.}{2013}]{2013MNRAS.434.2696S}
{Shahbaz} T.,  {Russell} D.~M.,  {Zurita} C.,  {Casares} J.,  {Corral-Santana}
  J.~M.,  {Dhillon} V.~S.,   {Marsh} T.~R.,  2013, \mn@doi [\mnras]
  {10.1093/mnras/stt1212}, \href
  {https://ui.adsabs.harvard.edu/abs/2013MNRAS.434.2696S} {434, 2696}

\bibitem[\protect\citeauthoryear{{Shemmer}, {Brandt}, {Netzer}, {Maiolino}  \&
  {Kaspi}}{{Shemmer} et~al.}{2008}]{2008ApJ...682...81S}
{Shemmer} O.,  {Brandt} W.~N.,  {Netzer} H.,  {Maiolino} R.,   {Kaspi} S.,
  2008, \mn@doi [\apj] {10.1086/588776}, \href
  {https://ui.adsabs.harvard.edu/abs/2008ApJ...682...81S} {682, 81}

\bibitem[\protect\citeauthoryear{{Strubbe} \& {Quataert}}{{Strubbe} \&
  {Quataert}}{2009}]{2009MNRAS.400.2070S}
{Strubbe} L.~E.,  {Quataert} E.,  2009, \mn@doi [\mnras]
  {10.1111/j.1365-2966.2009.15599.x}, \href
  {https://ui.adsabs.harvard.edu/abs/2009MNRAS.400.2070S} {400, 2070}

\bibitem[\protect\citeauthoryear{{Trakhtenbrot} et~al.,}{{Trakhtenbrot}
  et~al.}{2017}]{2017MNRAS.470..800T}
{Trakhtenbrot} B.,  et~al., 2017, \mn@doi [\mnras] {10.1093/mnras/stx1117},
  \href {https://ui.adsabs.harvard.edu/abs/2017MNRAS.470..800T} {470, 800}

\bibitem[\protect\citeauthoryear{{Tremou} et~al.,}{{Tremou}
  et~al.}{2018}]{2018ApJ...862...16T}
{Tremou} E.,  et~al., 2018, \mn@doi [\apj] {10.3847/1538-4357/aac9b9}, \href
  {https://ui.adsabs.harvard.edu/abs/2018ApJ...862...16T} {862, 16}

\bibitem[\protect\citeauthoryear{{Vasudevan} \& {Fabian}}{{Vasudevan} \&
  {Fabian}}{2007}]{2007MNRAS.381.1235V}
{Vasudevan} R.~V.,  {Fabian} A.~C.,  2007, \mn@doi [\mnras]
  {10.1111/j.1365-2966.2007.12328.x}, \href
  {https://ui.adsabs.harvard.edu/abs/2007MNRAS.381.1235V} {381, 1235}

\bibitem[\protect\citeauthoryear{{Vaughan}, {Edelson}, {Warwick}  \&
  {Uttley}}{{Vaughan} et~al.}{2003}]{2003MNRAS.345.1271V}
{Vaughan} S.,  {Edelson} R.,  {Warwick} R.~S.,   {Uttley} P.,  2003, \mn@doi
  [\mnras] {10.1046/j.1365-2966.2003.07042.x}, \href
  {https://ui.adsabs.harvard.edu/abs/2003MNRAS.345.1271V} {345, 1271}

\bibitem[\protect\citeauthoryear{{Wang}, {Wu}  \& {Kong}}{{Wang}
  et~al.}{2006}]{2006ApJ...645..890W}
{Wang} R.,  {Wu} X.-B.,   {Kong} M.-Z.,  2006, \mn@doi [\apj] {10.1086/504401},
  \href {https://ui.adsabs.harvard.edu/abs/2006ApJ...645..890W} {645, 890}

\bibitem[\protect\citeauthoryear{{Winter}, {Mushotzky}, {Reynolds}  \&
  {Tueller}}{{Winter} et~al.}{2009}]{2009ApJ...690.1322W}
{Winter} L.~M.,  {Mushotzky} R.~F.,  {Reynolds} C.~S.,   {Tueller} J.,  2009,
  \mn@doi [\apj] {10.1088/0004-637X/690/2/1322}, \href
  {https://ui.adsabs.harvard.edu/abs/2009ApJ...690.1322W} {690, 1322}

\bibitem[\protect\citeauthoryear{{Wrobel} \& {Nyland}}{{Wrobel} \&
  {Nyland}}{2020}]{2020ApJ...900..134W}
{Wrobel} J.~M.,  {Nyland} K.~E.,  2020, \mn@doi [\apj]
  {10.3847/1538-4357/aba8f7}, \href
  {https://ui.adsabs.harvard.edu/abs/2020ApJ...900..134W} {900, 134}

\bibitem[\protect\citeauthoryear{{Xie} \& {Yuan}}{{Xie} \&
  {Yuan}}{2017}]{2017ApJ...836..104X}
{Xie} F.-G.,  {Yuan} F.,  2017, \mn@doi [\apj] {10.3847/1538-4357/aa5b90},
  \href {https://ui.adsabs.harvard.edu/abs/2017ApJ...836..104X} {836, 104}

\bibitem[\protect\citeauthoryear{{Yang}, {Gurvits}, {Paragi}, {Frey}, {Conway},
  {Liu}  \& {Cui}}{{Yang} et~al.}{2020}]{2020MNRAS.495L..71Y}
{Yang} J.,  {Gurvits} L.~I.,  {Paragi} Z.,  {Frey} S.,  {Conway} J.~E.,  {Liu}
  X.,   {Cui} L.,  2020, \mn@doi [\mnras] {10.1093/mnrasl/slaa052}, \href
  {https://ui.adsabs.harvard.edu/abs/2020MNRAS.495L..71Y} {495, L71}

\bibitem[\protect\citeauthoryear{{Yesuf}, {Faber}, {Koo}, {Woo}, {Primack}  \&
  {Luo}}{{Yesuf} et~al.}{2020}]{2020ApJ...889...14Y}
{Yesuf} H.~M.,  {Faber} S.~M.,  {Koo} D.~C.,  {Woo} J.,  {Primack} J.~R.,
  {Luo} Y.,  2020, \mn@doi [\apj] {10.3847/1538-4357/ab5fe1}, \href
  {https://ui.adsabs.harvard.edu/abs/2020ApJ...889...14Y} {889, 14}

\bibitem[\protect\citeauthoryear{{Zhu}, {Brandt}, {Luo}, {Wu}, {Xue}  \&
  {Yang}}{{Zhu} et~al.}{2020}]{2020MNRAS.496..245Z}
{Zhu} S.~F.,  {Brandt} W.~N.,  {Luo} B.,  {Wu} J.,  {Xue} Y.~Q.,   {Yang} G.,
  2020, \mn@doi [\mnras] {10.1093/mnras/staa1411}, \href
  {https://ui.adsabs.harvard.edu/abs/2020MNRAS.496..245Z} {496, 245}

\bibitem[\protect\citeauthoryear{{den Brok} et~al.,}{{den Brok}
  et~al.}{2015}]{2015ApJ...809..101D}
{den Brok} M.,  et~al., 2015, \mn@doi [\apj] {10.1088/0004-637X/809/1/101},
  \href {https://ui.adsabs.harvard.edu/abs/2015ApJ...809..101D} {809, 101}

\makeatother
\end{thebibliography}

% Don't change these lines
\bsp	% typesetting comment
\label{lastpage}
\end{document}